# Minimax Is the Best Electoral System After All

Richard B. Darlington

Department of Psychology, Cornell University

## Abstract

When each voter rates or ranks several candidates for a single office, a strong Condorcet winner (SCW) is one who beats all others in two-way races. Among 21 electoral systems examined, 18 will sometimes make candidate X the winner even if thousands of voters would need to change their votes to make X a SCW while another candidate Y could become a SCW with only one such change. Analysis supports the intuitive conclusion that these 18 systems are unacceptable.

The well-known minimax system survives this test. It fails 10 others, but there are good reasons to ignore all 10. Minimax-T adds a new tie-breaker. It surpasses competing systems on a combination of simplicity, transparency, voter privacy, input flexibility, resistance to strategic voting, and rarity of ties. It allows write-ins, machine counting except for write-ins, voters who don't rate or rank every candidate, and tied ratings or ranks.

Eleven computer simulation studies used 6 different definitions (one at a time) of the "best" candidate, and found that minimax-T always soundly beat all other tested systems at picking that candidate. A new maximum-likelihood electoral system named CMO is the theoretically optimum system under reasonable conditions, but is too complex for use in real-world elections. In computer simulations, minimax and minimax-T nearly always pick the same winners as CMO.

# 1. Minimum change

## 1.1 Overview of Section 1

I argue that minimum change is the most important single property of an electoral system. Section 1.7 proposes a precise definition of minimum change, but for now I'll say merely that an electoral system has the minimum change property if it always selects as winner the candidate who would become the Condorcet winner with the least change to the voting results in some reasonable sense. Section 1.2 gives a 4-candidate example in which candidate D seems to me to be the obvious winner because of the minimum change principle, but Section 1.3 lists 18 well-known electoral systems which name D a loser.

Minimum change has obvious intuitive appeal, but its absolute indispensability is best conveyed by a discussion of the ways a Condorcet paradox (absence of a Condorcet winner) can arise. Section 1.4 presents a model of voter behavior, and Section 1.5 describes several ways in which that model allows a Condorcet paradox to arise. Section 1.6 argues that in each of these separate cases, a minimum-change approach is the most reasonable way to deal with the paradox.

The minimax, Dodgson (Felsenthal, 2012), and Young (1977) systems all meet reasonable criteria of minimum change, and all make D the winner by wide margins in the example of Section 1.2. Later sections argue that minimax is the best of these systems, for reasons including simplicity, transparency, voter privacy, and rarity of ties.

In minimax, election officials use the ballots already cast to "run" a two-way race between each pair of candidates. Minimax declares any Condorcet winner the winner. If there is none, minimax computes the largest loss (LL) for each candidate in their two-way races, and the candidate with the smallest LL is the winner. Minimax was suggested independently by Simpson (1969) and Kramer (1977). Section 3 describes three new tie-breakers, any of which can be added to classic minimax.

## 1.2 An example showing the importance of minimum change

I consider this example fairly realistic, given that it contains a Condorcet paradox, which any discussion of minimum change must contain.

A village received a state grant to establish a new park. Village officials identified three possible locations for the park. Spot A was heavily wooded, and was near a marsh, so it was inhabited by many bird species. Bird lovers in the community favored putting the park there. Spot B was slightly rolling and fairly open, but had many nice shade trees. This spot was favored by people who mostly wished to picnic in the new park. Spot C was treeless and very flat, and would provide an excellent site for both a soccer field and a softball field. The bird lovers were more attracted to the picnic site than the sports site, so their preference order was A B C. The picnickers liked sports more than bird-watching, so their preference order was B C A. The sports lovers preferred bird walks to sitting at picnic tables, so their preference order was C A B. The three groups were roughly equal in size.

It was understood that the next mayor would have a lot of say in choosing one of these sites. Three mayoral candidates all favored different sites. Call those candidates A, B, and C, according to the site they favored. This issue was the dominant issue in the mayoral election.

Also running was candidate D, who enjoyed bird-watching, picnicking, and sports about equally, and said that the choice of site needed further study. About half of the voters favoring each site thought that this was a reasonable position, given the divisions in the community, and ranked D first. All other voters felt strongly that a decision needed to be made promptly, and ranked D last. The bird lovers were evenly divided between voting patterns A B C D and D A B C, the picnickers were evenly divided between B C A D and D B C A, and the sports lovers were almost evenly divided between C A B D and D C A B. Specifically, 101 voters chose each of the first 5 patterns just mentioned, while 100 chose the last pattern. See Table 1.

**Table 1. Margins of victory in each pattern for the first-named candidate in each column heading**

| Pattern | Frequency | C>A | A>B | B>C | A>D | B>D | C>D |
|---|---|---|---|---|---|---|---|
| A B C D | 101 | -101 | 101 | 101 | 101 | 101 | 101 |
| D A B C | 101 | -101 | 101 | 101 | -101 | -101 | -101 |
| B C A D | 101 | 101 | -101 | 101 | 101 | 101 | 101 |
| D B C A | 101 | 101 | -101 | 101 | -101 | -101 | -101 |
| C A B D | 101 | 101 | 101 | -101 | 101 | 101 | 101 |
| D C A B | 100 | 100 | 100 | -100 | -100 | -100 | -100 |
| **Total** | **605** | **201** | **201** | **203** | **1** | **1** | **1** |

Table 1 reveals that every candidate loses at least one race. A loses to C by 201 votes, B loses to A by 201 votes, and C loses to B by 203 votes, while D loses to each of the others by just one vote. Thus D is the minimax winner "by a mile." The following points all support D as winner:

1. If just one more voter had decided that a prompt decision was unnecessary, and had therefore switched D from last place to first, D would become the Condorcet winner. Over 100 voters would need to change their votes to do this for any other candidate.

2. If two new voters had put D first, D would become the Condorcet winner. Over 200 such voters would be needed for any other candidate.

3. Those same two new voters would have given D an absolute majority of first-place votes. Over 400 such voters would be needed for any other candidate – about 4 times as many first-place votes as that candidate had actually received.

4. If two voters putting D last had failed to vote, D would become the Condorcet winner. Over 200 such voters would be needed for any other candidate. Thus D wins by the Young system, which counts the number of voters who must be removed to make each candidate a Condorcet winner.

5. The Dodgson electoral system (Felsenthal, 2012, p. 29) counts for each candidate the number of single-voter inversions of adjacent candidates needed to make that candidate the Condorcet winner; the candidate with the fewest inversions necessary is the Dodgson winner. In this example, moving D from last place to first for one voter requires 3 inversions of adjacent candidates for that voter. If we do that for any one of the 303 voters who had put D last, those 3 inversions will make D a strong Condorcet winner. At least 101 inversions would be required for any other candidate.

6. The new CMO system of Section 6 treats the observed data as a random sample from a larger population, as if each eligible voter had decided, randomly and independently, whether to vote in that election. CMO computes for each candidate X a likelihood ratio (LR) which measures the consistency between the data and the hypothesis that X is the Condorcet winner in that population. LR values can range from 0 to 1. We'll see in Section 6.4 that in the current example, LR for D is 0.9992 while the highest LR for any other candidate is under 1 in 600 trillion. Thus all four LR values are near theoretical limits. I argue in Section 6 that CMO is the theoretically optimum electoral system under reasonable assumptions. However, CMO is far too complex for use in most real-world elections. Simulation studies in Section 7 suggest that minimax picks the same winner as CMO in nearly all cases, just as it did in this example.

If we increased the 6 frequencies in Table 1 by equal amounts, even by thousands or millions, the largest margins of defeat for candidates A, B, and C would also rise by thousands or millions, while D's margins of defeat would all remain at 1. Thus in the 6-point list just presented, all the numerical differences between D and the other candidates can easily be increased without limit, making D an even more obvious winner.

### 1.3 How other electoral systems handle this example

As just described, the minimax, Young, Dodgson, and CMO electoral systems use what I consider reasonable measures of minimum change, and all name D as the winner in Section 1.2. The next paragraph describes a list I assembled, of what seem to be the best-known electoral systems aside from those just mentioned. All lack minimum change because they all name D a loser in this example.

Laslier (2012) asked 22 widely-recognized experts on electoral theory which of 18 electoral systems they "approved of." Each could approve as many systems as they wished. The number of approvals per system ranged from 0 to 15 with a median of 3.5. I included in my list all 12 systems which were approved by more than one of these 22 experts. Separately, Felsenthal (2012) discussed 18 electoral systems in detail. One of these ("successive elimination") is suitable only for choosing among versions of a legislative bill, not for human candidates. I included his 17 other systems. Another prominent author, Tideman (2006, p. 238), listed 22 electoral systems he considered worthy of discussion. Most are well known, but his list includes three little-known systems he considered inferior and two little-known systems he considered superior. I included all the well-known systems from his list, plus the two systems he considered superior. I merged the lists from these three authors, and eliminated duplicates. That yielded 21 systems – minimax, Young, Dodgson, and 18 others. In alphabetical order these 18 were: approval, Black, Borda, Bucklin, Coombs, Copeland, Hare (alternative vote), Kemeny (which Tideman calls "Condorcet"), majority judgment, Nanson, plurality, plurality with runoff, range, ranked pairs, Schulze, Schwartz, Tideman's alternative version of Schwartz, and Tideman's alternative version of Smith. I applied all 18 of these systems to the example of Section 1.2, and all 18 named D as a loser. A great many of these put D last of the four. Recall that D won overwhelmingly by all the various measures of minimum change mentioned above. It's therefore clear that except for minimax, Young, and Dodgson, all the best-known electoral systems violate the criterion of minimum change. Section 1.7 tells why I prefer minimax to the Dodgson system, and Sections 3.3 and 4.1 explain its advantages over Young, for reasons involving simplicity, transparency, and rarity of ties.

Because the principle of minimum change challenges the validity of so many well-known electoral systems, we should scrutinize that principle more closely. That scrutiny is facilitated by the model of voter behavior in the next subsection.

**1.4 A new model of voter behavior**

Tideman and Plassmann (2012) used the data from hundreds of real-world elections to test whether the assumptions of 12 electoral models fit that data. They concluded that 11 of the models fit poorly, and only one – a spatial model – fit well. In a pure spatial model, there is an axis for each relevant characteristic on which candidates vary, such as liberalism versus conservatism on economic, foreign, or social policies. Each candidate is represented by a dot in the space, and each voter is represented by a dot at that voter's ideal point. When any voter must choose between two candidates, the voter in the model always chooses the one closer to his or her own ideal point. The variables in a spatial model can be called "spatial" variables. Spatial models are sometimes described as allowing that voters may be "self-interested" rather than neutral "disinterested" observers. That's true, but the advantage of spatial models is greater than that. A rich person might feel that high taxes on the rich benefit the country as a whole by promoting economic equality. Or an ethnic majority member may support policies primarily benefiting ethnic minorities. Spatial models allow for differences like these among voters, without limiting the causes of the differences.

Tideman and Plassmann (2012) especially recommended low-dimensional spatial models. Following that advice, I used a two-dimensional model, but modified it in three ways to increase realism even further. In the new model, each voter has a certain level of *favorability* toward each candidate, and ranks the candidates in the order of their favorability values. Favorability is the sum of four or more values: (1) a value computed from spatial proximity as in a pure spatial model, (2) a value that might be called a candidate's "excellence" or "suitability" or "general attractiveness", (3) a value for each of one or more categorical variables, and (4) a random error term. Terms 2-4 are explained next.

Candidates have scores on excellence, but voters do not. The higher a candidate's score on excellence, the more attractive the candidate is to all voters. Excellence is a composite of traits like honesty, intelligence, health, articulate speech, generosity or selflessness, personal charm, experience in public office, and demonstrated heroism. Voters may not weight all these traits equally; some might weight honesty over heroism while others do the opposite. But each of these traits has an average weight across voters. Thus a weighted sum of these traits, using those average weights, provides a measure of excellence or general attractiveness. An excellence term in a computer simulation can consist simply of a normally distributed variable on which each candidate has some score.

Categorical variables are variables like religion, ethnicity, or preferred activity (as in the park-location example of Section 1.2). Each voter and each candidate falls in some category. For each such variable, the model includes a square table showing the average attitude of people in each category toward candidates in each category including their own. Values in such tables contribute toward each voter's assessment of each candidate. Of course, even the voters in a single category will not all be influenced in exactly the same way by a candidate's category membership. That's one reason for including the term described next.

The random error term subsumes several phenomena. One is the fact that a voter may misperceive a candidate's position on any of the other variables, thus randomly raising or lowering that voter's favorability toward that candidate. Misperceptions may be caused either by voter inattention or by deliberate deception by candidates, as when a candidate tries to appear one way to some voters and another way to others. The term also includes any effect specific to a particular voter-candidate combination. For instance, the voter and candidate might be next-door neighbors, and the voter would like to have such easy access to an elected official. In reference to the categorical variables of the previous paragraph, this term also allows that not all voters in a given category will be influenced in exactly the same way by a candidate's category membership.

I will not attempt to describe here the 11 non-spatial models which Tideman and Plassmann (2012) found to be unrealistic. In the terminology just introduced, those models all included excellence and error terms, but no spatial or categorical terms. The Borda (Felsenthal, 2012) and Kemeny (1959) electoral systems are perhaps the best-known systems based on models of this type.

In elections of most interest to electoral theorists, there are thousands or even millions of people who, in their fantasies, would like to win that election, but who have little or no demonstrated competence in positions even moderately like the one they seek. So there is a lot of variation on the excellence dimension. These people will also differ on the spatial and categorical traits. By the time some pre-election selection process has narrowed the number of candidates down from thousands to 2 or 5 or even 20, there has been severe selection on excellence, and much less selection on the other traits. Thus nearly all the relevant remaining variation is on these other traits. Therefore, a pure spatial model will fit the data far better than a model which has an excellence dimension but no spatial traits. That's what Tideman and Plassmann (2012) found.

## 1.5 Artifactual and non-artifactual paradoxes

Some instances of the Condorcet paradox seem to be artifactual, while others do not. In the former case, there is an important sense in which there is no paradox, even though one is observed. I argue in Section 1.6 that it is not extremely important to determine whether any particular instance of the paradox is artifactual, since minimax is the best electoral system for either case. But distinguishing between the two cases makes it easier to understand how paradoxes can arise.

The simplest non-artifactual paradoxes arise from categorical variables. In the park-location example of Section 1.2, ignoring candidate D reveals a large Condorcet paradox involving the other three candidates. For another example, suppose a city contains three economic or ethnic groups of about equal size. Everyone favors their own group. Group A tolerates B but detests C, while B tolerates C but detests A, and C tolerates A but detests B. In the mayoral election there is one candidate from each group. If everyone votes as predicted from these premises, a Condorcet paradox must occur. Both this example and the park example are obviously contrived, and my own guess is that non-artifactual Condorcet paradoxes arise only rarely. But we do need to include them in a list of possible cases.

Artifactual paradoxes are best explained in terms of spatial variables. A pure spatial model is said to have radial symmetry if voters are distributed symmetrically on any axis drawn through the center of the distribution. Plott (1967) showed that a Condorcet paradox can never occur in a population with radial symmetry, because in any two-way race, the candidate closer to the center of the distribution will win. But distance from the center is a transitive property, so if A beats B and B beats C, then A must also beat C. Univariate, bivariate, and multivariate normal distributions all have radial symmetry. But four types of distortion can produce Condorcet paradoxes even when radial symmetry exists in some deeper sense.

The first of these is voter carelessness and misinformation. If a computer simulation has generated voter ratings of candidates from a spatial model with a univariate, bivariate, or multivariate normal distribution, one can model increased carelessness and misinformation by adding mutually independent random errors to all ratings, thus introducing the error term of Section 1.4. These error values are mutually independent across both candidates and voters. Computer simulations (some of which appear in Section 5) show that adding such a term will often produce a Condorcet paradox when none existed before the error terms were added. One can model the upper limit of carelessness and misinformation by using simple random numbers as the voter ratings of candidates, making them independent across both voters and candidates. In simulations I have run using simple random numbers as voter ratings of candidates, the rate of paradoxes doesn't change greatly with the number of voters, but does increase with the number of candidates. For instance, with 75 voters, 40 candidates, and random ratings, I found 8082 paradoxes in 10,000 trials, but found "only" 2595 paradoxes in 10,000 trials with just 5 candidates. Thus under extreme levels of voter carelessness and misinformation, the rate of Condorcet paradoxes can be very high indeed, even with modest numbers of candidates. I have no evidence on the question, but I find it plausible that voter carelessness and misinformation may be the most common cause of the Condorcet paradox in real elections.

A second possible source of distortion is simple random sampling, as if every eligible voter decided, randomly and independently, whether to vote in that particular election. In that case, the votes cast would represent a random sample of a larger population of possible votes. If we use a pure spatial model with a univariate, bivariate, or multivariate normal distribution, the frequency of Condorcet paradoxes declines as sample size increases. Simulations of this type also appear in Section 5. Similar simulations show that with 10 candidates the paradox virtually disappears when sample sizes reach 10,000.

The third source of distortion is mathematically similar to the second, but we think about it in a different way. We all know that people's minds change over days, weeks, and even minutes. When opinion polls are taken every day for a week, it's not uncommon for the popularity of some candidate or policy to change from day to day, more than would be expected from random fluctuations in poll results. Thus Alice's vote cast on Election Day can be thought of as a sample of 1 from the population of the votes Alice might have cast over a period of a week or more. That way we can think of the election results as coming from a random sample of voter opinions even if the actual identities of the voters would change little or none from day to day. That means that if a Condorcet paradox appears on Election Day, it's still quite possible that there is no paradox in the larger population of votes which might have been cast at other nearby times.

The fourth type of distortion is asymmetric distributions of voter opinions. For instance, I have found that the rate of the paradox is noticeably higher when the spatial positions of voters and candidates are drawn from an asymmetric log-normal distribution (like the distribution of personal incomes) than from a symmetric normal distribution. Asymmetric distributions imply that the most extreme opinions at one end of an opinion dimension are noticeably closer to the median opinion than the most extreme positions at the other end.

## 1.6 Why minimum change is important for all Condorcet paradoxes

If a Condorcet paradox is caused by an artifact involving carelessness, misinformation, sampling error, or asymmetric distributions, then the paradox hides a "true winner" – a candidate who would have been the Condorcet winner except for the artifact. The artifacts that produced the paradox are usually of limited size, so that the "true winner" is likely to be the candidate who would become the winner with the least change in the data. Computer simulations in Section 5 support this conclusion.

Consider now a non-artifactual paradox caused by a categorical variable, as in the A-B-C examples of Sections 1.2 and 1.5. Once it has been announced that one candidate X has won under a Condorcet paradox, we can imagine a protest group forming for each of the candidates who had beaten X in two-way races. These groups all oppose each other as well as opposing X. Therefore, we want to minimize the size of the largest protest group. That goal is achieved by using minimax or some other minimum-change system.

Or suppose we have a categorical-variable paradox but it's reasonable to assume that people may change their minds over the course of several weeks or months. We'd like to be in a position so that only a few voters would have to change their minds to turn the announced winner into a Condorcet winner. That too suggests a minimum-change approach. If voters become used to some minimum-change electoral system based on preferential ballots, then pollsters may well start to use them, and to report poll results in terms of the margins of victory in several possible two-way races. Thus if a Condorcet paradox which existed on Election Day disappears a few months later in favor of the candidate who had been elected, people may well know it. That will make them more satisfied with the outcome, even if their own favored candidate hadn't won. Again, a minimum-change approach will increase the chance of that result. Or the election's winner may adjust his or her policies or alliances to try to reach a Condorcet-winner position in future polls, to assure re-election. The candidate who could do that most easily would be the minimum-change candidate. These arguments are very similar to one advanced by Young (1977, p. 350).

Thus a minimum-change approach seems best regardless of whether a Condorcet paradox is created by artifactual or non-artifactual causes.

**1.7 Why minimax uses the best measure of minimum change**

Because a Condorcet paradox may be produced in several different ways, it's very likely that no one electoral system will prove to be the mathematically perfect solution for all cases. We will see in Section 6 that CMO has a good claim to mathematical perfection for the random sampling cases mentioned earlier, though even there we must use the questionable assumption of mutually independent voters. But I will argue that we can narrow down the choice to several very similar systems, and then choose among those systems using the criteria discussed in Sections 3 and 4.

The Copeland electoral system might appear on the surface to be a minimum-change approach. In the Copeland system, the candidate who wins the most two-way races is the winner, so he or she is the one who would have to win the fewest additional races to become a Condorcet winner. However, the park-location example of Section 1.2 shows that the Copeland system can conflict with more appealing measures of minimum change. In that example, D loses all his or her two-way races, while A, B, and C lose only one race each. But only 1 voter would have to change their vote to make D a Condorcet winner, whereas over 100 changes would be required for any other candidate. People naturally count voters rather than candidates when assessing election outcomes. If a plurality (vote for one) election has 10 candidates and the top two candidates get 2000 and 1000 votes respectively, we would never say that the race between them had been close because the top candidate had beaten only one more candidate than the second-place finisher had beaten.  I thus reject Copeland as a reasonable minimum-change method.

The Dodgson system counts the number of single-voter switches between adjacent candidates needed to turn a given candidate into a Condorcet winner. However, I consider this the weakest rule which still qualifies as a minimum-change rule. Three switches are needed to change A B C D to either D A B C or to C B A D. But suppose A, B, and C are very similar (e.g., very liberal) while D is very different from them all (e.g., very conservative). Then the switch from A B C D to D A B C implies a complete change in a voter's orientation, from liberal to conservative, while the switch to C B A D implies no real change in orientation. Despite some definitions of "clones," no proposed electoral rule includes any general measure of the sizes of similarities or differences among candidates, and it would be completely impractical to try to include one. Therefore, I consider the Dodgson rule to be the least useful of the several measures which I still consider to be minimum-change measures.

Three other measures of minimum change count the number of voters who must change in some way to make each candidate X a Condorcet winner, and name as winner the candidate for whom this number is smallest. One way is to count the number of voters who would need to change their votes to put X first. This possibility is not in Section 1.3's list of 21 well-known systems, presumably because of its potential computational complexity. A second way was suggested by Young (1977); we count the number of voters we would need to delete who had ranked X poorly. The third way is to count the number of new voters we would have to add if they all put X first. The last of these is equivalent to minimax, since each new voter putting X first would reduce each of X's margins of defeat by 1. Young (1977) mentioned (p. 350) that his system and minimax are probably very similar in results. It seems clear that the opinion-change method is also very similar, since it's effectively a combination of the other two. That is, changing a voter's vote from anti-X to pro-X is equivalent to deleting one anti-X voter and adding a pro-X one. But of these three similar systems, minimax has two advantages. First, we will see in Section 3 that minimax has a simple tie-breaker not readily available in the other systems. Second, we will see in Section 4.1 that minimax surpasses the other two methods on a combination of simplicity, transparency, and voter privacy. Section 3 describes several new forms and relatives of minimax which have these same advantages.

## 2. Ten electoral criteria conflicting with minimum change

Section 1 concluded that any acceptable electoral system must satisfy minimum change, and therefore must also be Condorcet-consistent (picking every Condorcet winner as the winner). That in turn implies the necessity of majority rule in two-candidate elections. This section examines 10 well-known and plausible-seeming electoral criteria which must be discarded because they conflict with minimum change, Condorcet consistency, or majority rule. Many of these 10 present other problems as well. I believe these 10 criteria have been accepted by some because most people will accept a criterion as reasonable if: (a) ordinary plurality (vote for one) voting meets the criterion, and (b) they cannot easily imagine why any plausible system would fail to meet the criterion. This approach leads to unfortunate results because most people cannot imagine all the potential consequences of a Condorcet paradox. When these consequences are illustrated, it becomes easier to see why these criteria must be abandoned.

One prominent electoral theorist who appears to agree with me on many of these issues is Tideman (2006). In Table 13.2 on page 238 he divides 22 electoral systems into 5 tiers of "supportability." Minimax is one of just five systems in his highest tier, despite the fact that most of the criticisms of minimax described below have been well known for decades. Although I join Tideman in rejecting these criticisms, I differ from him within his highest tier, because the other four systems in that tier all fail my criterion of minimum change. These systems are ranked pairs, Schulze, "alternative Schwartz," and "alternative Smith."

The 10 criteria in question are:

1. The **Condorcet loser** criterion states that no candidate should win an election if they lose all their two-way races. Thus it directs that in the example of Section 1.2, D must lose.

2. The **absolute loser** criterion states that no candidate should win an election if they are ranked last by over half the voters. Thus in Section 1.2, D must lose.

3. The **preference-inversion** criterion (Saari, 1994) states that an electoral system is unacceptable if it names candidate X the winner, but would still name X the winner if all voter rankings of candidates were inverted. If rankings were inverted, candidates A, B, and C in Section 1.2 would still each lose a two-way race by over 200 votes while D would become the Condorcet winner and thus the minimax winner, beating every other candidate by one vote. Therefore, any method which named D the original winner violates the preference-inversion criterion.

4. The **mutual majority** criterion says that if there is a set of candidates such that more than half the voters prefer every candidate in the set to everyone outside it, then some candidate in the set must win. In the example of Section 1.2, Candidates A, B, and C form such a set, so D must lose.

5. The **Smith** criterion says that every winner must come from the Smith set, which is the smallest set of candidates such that everyone in the set beats everyone outside the set in two-way races. In the example of Section 1.2, the Smith set includes A, B, and C, so D must lose.

6. The criterion of **independence of irrelevant alternatives (IIA)** states that when deciding between any two candidates A and B, the only relevant information is the voters' responses concerning those two candidates. IIA is one of the criteria which Tideman (2006) explicitly rejects; he says that the arguments for IIA are "not convincing" (p. 132). Here I'll offer my own reasons for rejecting IIA.

IIA implies that an election's winner should not change if votes are recounted after one of the losers drops out. But in the example of Section 1.2, D would lose if any of the other candidates dropped out, since the Condorcet paradox would no longer exist. Thus any electoral system violates IIA if it selects D in that example.

IIA conflicts with the majority-rule criterion whenever there is a Condorcet paradox with three candidates. No matter which candidate any system picks, that candidate loses to some other candidate in a two-way race. Therefore, if those two stay in the race and the other candidate drops out, the system would violate majority rule if it stayed with the same winner as required by IIA.

Even in the absence of a Condorcet paradox, the very phrase "irrelevant alternatives" confuses a losing candidate (the "irrelevant alternative") with the data that was collected because that candidate was in the race. That data may be useful in choosing among other candidates even though the one candidate lost. For instance, suppose you planned to bet on a forthcoming game between teams A and B. Those teams had played each other only once, and A had lost to B by 1 point, 13 to 14. But both teams had recently played team C, and A had beaten C by 10 points while C had beaten B by 10 points. C is "irrelevant" to the forthcoming game, since it's not even playing. But the data involving C seems highly relevant in betting on the forthcoming game. Similarly, suppose that in a 10-candidate race, A beats B by 1 vote, but B beats all the other 8 candidates by much broader margins than A beats them. It's not obvious that the latter fact should be ignored as IIA requires. Indeed, a computer simulation study in Section 3.2 found that an electoral system conforming to IIA was, on the average, less good at picking the best winners than an alternative system which violates IIA.

IIA's conflict with majority rule means that nearly all electoral systems violate IIA, since they reduce to majority rule in two-candidate elections. The best-known electoral system satisfying IIA is the majority judgment system of Balinski and Laraki (2007, 2010). It is hard to see how one could accept IIA without accepting this system. But this system has several problems of its own. Table 3.2 in Felsenthal (2012) lists 10 faults with majority judgment, three of which Felsenthal rates as major. This is the worst total attained by any of the 18 systems rated by Felsenthal in his Tables 3.1, 3.2, and 3.3. All this suggests that IIA can be dismissed.

7. The criterion of "**independence of clones**" says that if candidate X would lose a two-way race to any one of a group of "clones," then X should not win when all those clones are in the same race against X. This criterion typically uses the definition of clones accepted by Tideman (1987) and others, which defines clones as a group of candidates so similar that no candidate outside the group is ranked between any two of the clones by any voter, or tied with any of the clones. In the example of Section 1.2, A, B, and C are clones by this definition, and D loses to all of them, so D most lose.

I feel that the definition of "clones" just mentioned is actually unreasonable. That word implies to most people that the candidates referred to are perceived as very similar by all voters. But if two candidates are genuinely perceived that way, we would not observe one of them beating the other 2:1 in a two-way race, as A, B, and C all do in the current example. When Tideman (1987) first charged that minimax lacks independence of clones, the between-clone percentage margins of victory in his example (p. 195) were also very large, at 19%, 33%, and 48% of voters, and that was an essential feature of his example as it is in mine. A more reasonable definition of "clones" would require also that the margin of victory between any two clones must be smaller than any margin between a clone and a non-clone. Under that definition, neither Tideman's example nor mine would have any clones. Thus systems like minimax, which satisfy the minimum-change criterion, would no longer violate the independence-of-clones criterion.

All of the seven criteria listed above are violated by any electoral system which chooses candidate D in the example of Section 1.2. Thus those criteria must be rejected by anyone who feels that D should win in that example. Other reasons were given for rejecting some of these criteria, but the example of Section 1.2 spans all of them. The next two criteria in my 10-item list produce anomalous results in other examples, as shown below.

8. The **participation** criterion states that if candidate A wins an election, and we then add more voters who all prefer A to B, the new votes should never change the winner to B. But Moulin (1988) showed that any Condorcet-consistent system must violate this criterion. Table 2 gives an example in which I feel minimax makes a reasonable choice even though that very example illustrates minimax's failure to satisfy the participation criterion. This table says for instance that two voters put A first, then B, then D, then C; six voters put B first, etc.

Like all examples purporting to show minimax anomalies, this example contains a Condorcet paradox: A beats B, B beats C, but C beats A. Working through Table 2 reveals that in the six two-way races, the largest losses for candidates A, B, C, and D, are by margins of 10, 4, 6, and 10 votes respectively. Thus B is the minimax winner, since it has the smallest value in this list. Suppose we then add two more voters, both with pattern A B C D. The largest losses are now 8, 6, 4, 12, in the same order, so C is now the minimax winner. Thus the two additional voters changed the minimax winner from B to C, even though both new voters had placed B before C.

**Table 2. Frequencies of 5 voting patterns**

| Freq. | Pattern |
| --- | --- |
| 2 | A B D C |
| 6 | B D C A |
| 5 | C A B D |
| 1 | D A B C |
| 2 | D C A B |

This puzzling result occurred because with the 16 original voters, B's largest loss was to A whereas C's largest loss was to D. But the two new voters ranked A first and D last, thus strengthening A and weakening D relative to B and C. This increased B's loss to A and decreased C's loss to D, just enough to change the relative sizes of those two losses, and thus change the minimax winner. Thus when we consider the total voting pattern of the two new voters, and not just the fact that they preferred B to C, the minimax result seems reasonable.

9. The **consistency** criterion states that an electoral system is unacceptable if it's possible for candidate X to win in each of two sets of votes (e.g., in two districts), but lose when the two sets are merged into one. But Young and Levenglick (1978) show that every Condorcet-consistent system except Kemeny (1959) must violate that criterion, and Section 1.2 shows that Kemeny violates the minimum-change criterion. Table 3 presents an example, found on the internet, which shows that minimax violates the consistency criterion. But I will argue that even in this example, the minimax choice is the most reasonable one.

**Table 3. Frequencies of 8 voting patterns**

| Freq. | Pattern |
|-------|---------|
| 1 | A B C D |
| 6 | A D B C |
| 5 | B C D A |
| 6 | C D B A |
| -------------------- | |
| 8 | A B D C |
| 2 | A D C B |
| 9 | C B D A |
| 6 | D C B A |

When we consider just the first four voting patterns in Table 3, the largest margins of defeat for A, B, C, and D, are respectively 4, 6, 6, and 6, so A is the minimax winner. Considering just the last four patterns in the table, the largest margins of defeat are respectively 5, 9, 7, and 9, so A wins again. But when all eight patterns are analyzed together, the largest margins are respectively 9, 3, 1, and 3, so C is now the minimax winner. The feature producing this odd result is that B, C and D form a Condorcet cycle in the first group, with all within-cycle margins of defeat exceeding any involving A. That happens again in the second group. But in the second group the cycle runs in the opposite direction. That is, in the first group, B beats C, C beats D, and D beats B, but all those margins are reversed in the second group. When the two groups are combined, the two cycles largely cancel each other out. There is still a cycle among those candidates, but all the margins of defeat are now much smaller than in the previous cycles. All A's margins of defeat are in the same direction in the two portions, so they supplement each other instead of canceling out. Therefore, A's margins grow to exceed any of the now-smaller within-cycle margins, making A lose. We might say that each half of the data appeared to give a reason why A should beat C. But in each case the other half of the data contradicted the conclusion from the previous half, thus making it clear that A should lose.

10. The **truncation** criterion says that a voter should never derive any benefit from deliberately recording ranks or ratings only for his or her most-favored candidates rather than for all candidates. However, Felsenthal (2012) notes that this criterion leads to the dismissal of not only all 8 of the Condorcet-consistent electoral systems he discusses, but of all 14 systems he discusses which ask voters to rank the candidates. Thus the truncation criterion is clearly too severe. Many examples of the truncation paradox assume unrealistically that the truncating voter knows exactly how all other voters voted or will vote, and uses that information to benefit himself or herself.

Thus there are good reasons to discard all 10 of the criteria in this section. That leaves minimax free of all the major faults which have in recent decades led to its rejection.

## 3. Some new forms and relatives of minimax

### 3.1 Ballot format and some terms

As already mentioned, minimax (and most of the systems we discuss) use ballots which allow a voter to rank all candidates, or to rate them on a scale with at least 3 points. Such ballots are typically called **preferential ballots.** Most of these systems can be used with many different ballot formats, but the sample ballot shown below is machine-readable except for ballots with write-ins.

---

**SAMPLE BALLOT**

**Voting directions**. In the boxes on the lower right, you may write in the names of as many as 3 candidates not listed on the ballot. You may express your evaluation of each listed candidate and each write-in by filling in one circle in the appropriate row of circles. The greater your preference for a candidate, the further to the left your mark should be. You may express equal preference for two or more candidates. If you leave blank all the circles for one candidate, that candidate will be ranked below all the candidates for whom you did fill in circles. If you fill in two or more circles on a single row, only the first circle will be counted. The system counts only how you rank the candidates, so changes in your marks which don't change the ranking will not change your vote.

| Allen Able | ○ ○ ○ ○ ○ ○ ○ | Write-in #1 |
| Betty Barton | ○ ○ ○ ○ ○ ○ ○ | |
| Charles Carr | ○ ○ ○ ○ ○ ○ ○ | |
| Delores Drew | ○ ○ ○ ○ ○ ○ ○ | Write-in #2 |
| Write-in #1 | ○ ○ ○ ○ ○ ○ ○ | |
| Write-in #2 | ○ ○ ○ ○ ○ ○ ○ | |
| Write-in #3 | ○ ○ ○ ○ ○ ○ ○ | Write-in #3 |

---

Optionally, each column of circles may be topped by an evaluative term like "good" or "poor," or by a number or letter (in order starting with 1 or with A); those options are not illustrated here. To eliminate any confusion, voters may be given the option of studying these directions, and a sample ballot, in newspapers before Election Day, and children may study such ballots in school.

The one minor disadvantage of this format is that it limits the number of write-ins any one voter may list. Election officials bothered by this can always increase the number of write-in boxes. If each row of circles has at least as many circles as there are listed candidates plus allowed write-ins, the voter is free to give a complete ranking of all the candidates, but is not forced to do so.

If some voters give two or more candidates the same rating, or write in names, or fail to rate all candidates, we'll say we have **partial ranking**. Votes with none of these complications will be said to exhibit **full ranking.** Historically, some real-world elections using preferential ballots have demanded full ranking. But many electoral systems, including minimax-T, can be employed with either full or partial ranking, and with or without verbal or other column headings.

If a voter fails to rate or rank some candidate, we will consider that voter to have placed that candidate below all others whom they did explicitly rate or rank. That seems reasonable because the voter who omits candidate X is saying essentially that they're so uninterested in X that they didn't even take the time to form a clear opinion about him or her. If a voter puts a mark for X to the right of all other marks on that voter's ballot, the system will record that the voter prefers all other listed candidates to X but prefers X to all write-ins by other voters whom the current voter didn't also write in. But if a voter makes no mark at all for X, the system will record X as equal to all those write-in candidates. Thus the strongest way to vote *against* X is to put no mark next to X.

Preferential ballots allow election officials to "run" a two-way race between each pair of candidates. A candidate who wins all their two-way races is called a **strong Condorcet winner**. I'll call a candidate a **weak Condorcet winner** if they at least tie all their two-way races, provided every other candidate loses at least one race. That result is actually possible, and would indeed occur if one more voter were added to the last pattern in Table 1. An electoral system is **Condorcet-consistent** if it picks every Condorcet winner as winner. Several well-known electoral systems are not Condorcet-consistent; see Table 4. A set of candidates with no Condorcet winner, as when candidate A beats B, B beats C, and C beats A, forms a **Condorcet cycle**, and the situation itself is called a **Condorcet paradox**. The simplest possible Condorcet paradox occurs when one voter puts three candidates in the order A B C, a second voter records B C A, and a third records C A B. Then A beats B 2:1, and B beats C 2:1, but C beats A 2:1.

### 3.2 Some minimax variants and tie-breakers

Ties can be a severe practical problem, especially with fewer voters. The simulation studies of Section 5 show that several electoral systems produce ties in a very noticeable fraction of all trials exhibiting a Condorcet paradox. Examination of other systems (not studied in Section 5) suggests that they too could be very subject to this problem. Minimax suffers from this problem. In one full-ranking simulation with 4 candidates and 75 voters in each of 1000 trials with Condorcet paradoxes, the number of ties in classic minimax was 386. That's clearly unsatisfactory, and the number of ties is even larger with fewer voters per trial; with only 35 voters per trial, the number of ties was 544 out of 1000. This section describes three new tie-breakers I'll call minimax-T1, -T2, and -T3, in order of increasing complexity. The T stands for "tie-breaker." In earlier sections I used the term minimax-T to stand for this set of three methods collectively. All three of these methods are simple enough for an ordinary voter to check the calculations, thus eliminating suspicions of fraud at this step. However, the later methods require more steps.

Two new variants of minimax, which I'll name minimax-P and minimax-Z, may produce far fewer ties than classic minimax under partial ranking. Define the number of "participants" in each two-way race as the number of voters who expressed some preference between the two candidates in that race. Minimax-P expresses the margin of victory in each two-way race as a proportion of the number of participants in that particular race, and uses these proportional margins the same way classic minimax uses the raw margins. Thus in the event of a Condorcet paradox, the winner is the one whose largest proportional loss is smallest.

To understand minimax-Z it helps to think of the participants in each two-way race as a random sample from a larger population. The sign test is the form of the binomial test which tests the null hypothesis that two non-overlapping frequencies are equal. Minimax-Z draws on the normal approximation to the sign test. As applied to a two-way race, this test uses the formula $z = (W\text{-}L)/\sqrt{(W+L)}$, where $W$ and $L$ denote respectively the winning and losing frequencies in the race. In minimax-Z we take each candidate's largest absolute loss and convert it to $z$. The winner is the candidate with smallest $z$. The comparable expressions for classic minimax and minimax-P are respectively ($W$-$L$) and ($W$-$L$)/($W$+$L$).

Could minimax-P or minimax-Z replace minimax? I ran a simulation study comparing these three methods on their ability to pick the most centrist candidate – the candidate closest to the center of the array of voters. For simplicity I defined the "center" in terms of means rather than medians; I assume that decision had little effect on the study's results. On each trial the computer generated 75 voters and 5 candidates in a bivariate standard normal spatial model. Voter ratings of candidates were rounded to the nearest integer on a 9-point scale, so some of these computer-generated "voters" might give the same ratings to two or candidates. Thus this simulation used partial ranking. The computer had to generate over 12 million trials to find 20,000 trials exhibiting a Condorcet paradox. Classic minimax, minimax-P, and minimax-Z were used to pick winners in these 20,000 trials. Classic minimax produced ties in 8107 of these 20,000 trials, minimax-P in 875 of the trials, and minimax-Z in 841 of them. Thus in this study, minimax-P and minimax-Z both produced fewer than 1/9 as many ties as classic minimax.

But further analysis of these 20,000 trials suggests that classic minimax is better than minimax-P, and about equal to minimax-Z, at picking the most centrist candidates when there are no ties. In the overwhelming majority of the tie-free trials, the three methods all picked the same winners. But I compared each pair of methods on the trials in which those two methods showed no ties and also picked different winners. I compared each pair of methods on the centrism of the winner they picked. Centrism values were computed to 16 digits and were not rounded, so no two candidates were ever exactly tied on centrism. In these comparisons, classic minimax picked a more centrist candidate than minimax-P in 106 trials and less centrist in 64 trials. Minimax-Z picked a more centrist candidate than minimax-P in 91 trials and less centrist in 58 trials. Classic minimax picked a more centrist candidate than minimax-Z on 15 trials and less centrist on 8 trials. The first two of these three differences (the ones involving minimax-P) were statistically significant beyond the .01 level by the sign test, while the difference between classic and minimax-Z was not even close to significant.

Thus when there are no ties, classic minimax seems the best of the three. It's significantly superior to minimax-P at selecting the most centrist candidate, and it's noticeably simpler than minimax-Z. Of course, simplicity is very important in public elections. That suggests using classic minimax as the basic method, and using minimax-P (which is simpler than Z) as a tie-breaker. I'll call that procedure minimax-T1.

The study just described is the first of four simulation studies in this section. All four are broadly similar in design, differing only in the ways mentioned below.

A very different tie-breaker is possible. I'll call it minimax-H, where the H stands for "head to head." If just two candidates are tied for winner in classic minimax, the minimax-H winner is the one who beat the other in their two-way race. If there are three or more tied candidates, we look for a Condorcet winner among just those candidates, and that candidate is the final winner.

I compared minimax-H to minimax-T1. Using a bivariate normal spatial model with 5 candidates and 75 voters in each trial, I generated enough trials so that there were 10,000 Condorcet paradoxes with two or more candidates tied for winner by classic minimax. The study was a bit oversimplified but still seems fair in the way it compared the two methods under study: if three or more candidates were tied, the computer program examined just the first two of the tied candidates. A tie-breaker was considered to have "hit" if it picked the more centrist of the two candidates being compared, and to have "missed" if it picked the less centrist one. Minimax-T1 scored 5344 hits, 424 ties, and 4232 misses, thus scoring more hits than ties and misses combined. Minimax-H scored 4229 hits, 0 ties, and 5771 misses, thus scoring more misses than hits. Thus minimax-T1 seems clearly superior to minimax-H as a tie-breaker. Incidentally, this conclusion contradicts the IIA criterion (independence of irrelevant alternatives), since minimax-H is consistent with IIA and minimax-P is not. I already rejected IIA in point 6 of Section 2.

One might wonder how minimax-H could possibly score far more misses than hits. Psychologists recognize a phenomenon they call "suppression" of sources of error in psychological tests and measures. Suppose we have two tests of knowledge of history. Both tests are speeded, meaning many students cannot finish the tests in the allotted time. Both tests produce the same mean scores, and both have the same range in large groups, with scores ranging from about 20 to 80. But test B is much more speeded than test A. Suppose Frank and Clare achieve the same score on test A, but Frank outscores Clare on test B. That suggests that Frank can work faster than Clare on tests like these, since Frank's margin over Clare was greater on the more speeded test. But since test A was also speeded, and Clare did as well as Frank on test A even though Frank worked faster, that suggests that Clare actually knows more history. Thus test B actually has negative value when breaking ties on test A, giving higher scores to those who know less history, even though B would have positive value if test A were unavailable. Something like this appears to be occurring with minimax-H. It apparently has negative value as a tie-breaker for classic minimax, even though H would obviously have positive value if it had to be used alone. For readers familiar with multiple regression, a suppressor variable X is defined more generally as a variable which receives a significantly negative weight in multiple regressions despite correlating positively with the variable being predicted. If this occurs, it is presumed that X is measuring the sources of error in the other predictor variables (reading speed in the current example), so that X's negative weight subtracts out or suppresses those sources of error. Suppression is discussed in many texts on multiple regression, including Darlington and Hayes (2017), pp. 200-201.

Thus I recommend minimax-T1 over minimax-H as a tie-breaker for classic minimax. This is the simplest of the three tie-breakers I consider reasonable.

Minimax-T1 will not break any ties under full ranking, since it will break a tie between the winners of two two-way races only if those races have different numbers of participants, and full ranking means all voters participate in all two-way races. Minimax-T2 can break ties under either full or partial ranking. Even under partial ranking it may break more ties than minimax-T1, especially if there are many candidates and few voters. In minimax-T2 we think of each margin of defeat as a negative margin of victory. Consider all of candidate X's margins of victory, expressed as some mixture of positive and negative numbers. In minimax-T2 we put those numbers in a column and sort them from most negative to most positive. We needn't keep track of the opponents against whom these victories or defeats were scored; we record merely the margins themselves. We do this for each candidate. Then in simple minimax the winner is the candidate in whose column the first entry is least negative. If two or more candidates are tied on these first column entries, minimax-T2 compares them on their second column entries. If any of those candidates are tied on those values, it compares them on their third entries, and so on. When a tie-free set is found, the winner is the candidate whose margin is least negative or most positive.

In 100,000 trials with Condorcet paradoxes under full ranking, each with 4 candidates and 75 voters, minimax-T2 left only 679 ties. It thus cut the rate of ties by a factor of 57 relative to the first study in the opening paragraph of this subsection, since (386/1000)/(679/100,000) = 57. Since minimax-T1 reduced the number of ties by a factor of about 9 in a different study, this crude comparison between them suggests that minimax-T2 breaks even more ties than T1. The greater the number of candidates, the fewer ties minimax-T2 produces. In a set of studies in Section 5 which had 10 candidates and 75 voters in each trial, only 2 ties were observed in 300,000 trials with Condorcet paradoxes.

Since minimax-H actually does worse than chance in breaking ties, we might ask whether the same is true of minimax-T2. I studied whether minimax-T2 performs better than arbitrarily picking a winner when there is a tie. Using a spatial model with 10 candidates and 75 voters, I compared minimax-T2 to the arbitrary choice on the centrism of the candidates they chose. Both voters and candidates were drawn anew for each trial. In 1 million trials, all displaying a Condorcet paradox, using classic minimax, two or more candidates tied for winner in 10,487 trials. In 5001 of those 10,487 trials, the arbitrary tie-breaker and minimax-T2 picked the same winner. In the remaining 5486 trials, minimax-T2 picked a more centrist candidate than the arbitrary choice in 3468 trials, and less centrist in 2018 trials. So we can conclude that in this small but informative fraction of trials, minimax-T2 does substantially outperform an arbitrary choice at picking the more centrist candidate. Of course a non-arbitrary system like minimax-T2 would also be more satisfying to voters than an arbitrary choice.

Minimax-T3 combines minimax-T1 and minimax-T2. It can thus improve on T2 only under partial ranking, since T1 breaks no ties under full ranking. First minimax-T3 applies T1, which uses the proportional form of what (in the terminology of minimax-T2) is each candidate's most-negative raw margin of victory. If this step doesn't break the tie, T3 examines each candidate's second-most-negative raw margin of victory. The winner is then the candidate for whom this number is least negative or most positive. If a tie still remains, the analyst converts these second-most-negative raw margins to proportional margins. Again, the winner is the one for whom this value is least negative or most positive. If a tie still exists, proceed to each candidate's third-most-negative raw margin of victory, and so on. Thus minimax-T3 alternates between comparing raw margins and proportional margins. It is more complex than T1 or T2, but should break nearly all ties under partial ranking. I performed no separate simulations to test T3's ability to select the most centrist of the tied candidates. However, T3 is just a combination of T1 and T2, and simulation studies did show that both those methods outperform chance in their selection of winners.

In summary, the three recommended tie-breakers of this section are simple and transparent, thus avoiding suspicions of fraud, they are natural extensions of classic minimax, some will break nearly all ties, and simulations suggest that they will select winners who on the average are more centrist than winners chosen by arbitrary tie-breakers.

### 3.3 SSMD and SSSMD

I created and studied two other extensions of minimax. SSMD stands for "smallest sum of margins of defeat." In this system, we sum all of candidate X's margins of defeat in two-way races (ignoring X's victories in those races), and the winner is the candidate with the smallest sum. SSSMD stands for "smallest sum of squared margins of defeat." The method's rule should be obvious from its name. After creating these methods, I learned that SSMD is the same method that Tideman (2006, p.

199) called the simplified Dodgson method. Neither SSMD nor SSSMD performed nearly as well as minimax-T2 in the simulations of Section 5. SSMD and SSSMD seem less closely related to minimax than the other methods of Section 3, so I'll call them "relatives" rather than "forms" of minimax.

## 4. Process Criteria for Electoral Systems

The "product" of an electoral system is the chosen winner. Previous sections focused on that. This section discusses the "processes" of various systems.

### 4.1 Simplicity, transparency, and voter privacy

An electoral system's simplicity refers to the ease with which a typical voter can understand how to vote and how an election's winner is determined. Transparency refers to the ease with which a voter can personally see and check the steps by which a winner is determined, to avoid suspicions of fraud by insiders. Simplicity and transparency are especially important in public elections, where voter sophistication is low, stakes are high, and fraud is historically common.

Naish (2013) describes a conflict between transparency and voter privacy. Transparency has sometimes been maximized by publishing replicas of all ballots, to allow anyone to literally count the votes themselves. According to Naish, it is believed that the Italian mafia has taken advantage of this practice to bribe or coerce voters. A corrupt outsider can direct a voter to put the outsider's favored candidate first and to use the lower-ranked ballot choices to record an identifiable "signature." For instance, this might consist of an alternating pattern of far-left and far-right candidates, of a sort that would rarely if ever appear naturally. Because of such concerns, I will assume that the maximum feasible level of transparency is achieved by publishing merely all raw vote totals, so that anyone who wishes can check how those totals were analyzed to select a winner. In classic minimax, for instance, that checking process would include computing all two-way margins of victory and defeat, identifying each candidate's largest margin of defeat, and finding the smallest of those – all steps that many voters could easily repeat for themselves, starting from the vote totals.

Simplicity is a necessary but not sufficient condition for transparency. For instance, in the classic Hare system, the candidate with the fewest first-place votes is eliminated and votes are then recounted. This is repeated until only one candidate is left. That's easy to understand, but checking the calculations requires repeated access to every single ballot, so transparency in Hare must be sacrificed to protect voter privacy. Several other systems have the same problem, including Young (1977) and Dodgson. Some other systems protect voter privacy because the only data they use to find the winner are the vote totals of two-way races, but have calculations so complex that the average voter cannot repeat them, thus destroying transparency. The Schulze (2011) and Tideman (1987) systems are two such systems. Section 1.7 mentioned the possibility of counting the number of new voters, all putting X first, who would have to be added to make X the Condorcet winner. That system combines simplicity, transparency, and voter privacy. As already mentioned, it is equivalent to minimax.

All the aforementioned forms and relatives of minimax combine voter privacy with reasonable levels of simplicity and transparency. Specifically, if a 15-year-old student is given a table showing the vote totals in all two-way races, for any of these methods they should be able to determine the winner using only pencil, paper, a four-function calculator, and just a few minutes, without having to read the method's printed directions more than once. Just four well-known competing systems can meet this standard. One is ordinary plurality (vote for one) voting. Another is approval voting, in which each voter "approves" as many candidates as they like, and the candidate with the most approvals wins. Two others are the Copeland and Borda systems. In Borda, candidate X's "score" from each voter is the number of other candidates the voter ranks below X. The winner is the candidate whose scores have the highest sum. However, these four systems all have other disadvantages, as described in the next few subsections.

## 4.2 Input flexibility

Preferential ballots allow voters to respond in many different ways. A voter may give a strict ranking of all candidates named on the ballot, or rank only his or her top three of 10, or name only one most-preferred candidate, or rate candidates on two-point scales or three-point scales, etc. A voter who wishes to respond in a simple manner, for instance by naming only one candidate, is not discomfited by the fact that the ballot format allows more complex patterns of response. I'll call this *input flexibility*. Many electoral systems use preferential ballots and thus score high on input flexibility. Plurality voting and approval voting are the only systems mentioned here with little input flexibility. That can lose precision and add frustration, because voters are not allowed to express all their preferences. Simulation studies in Section 5 show that when one uses any of three different ways to define a "best" candidate in certain artificial situations, both plurality and approval voting are much less likely than most other systems to pick that "best" candidate. I presume this is because these systems do not use nearly all the information which voters would express on preferential ballots.

## 4.3 Susceptibility to strategic voting

Another problem with plurality and approval voting is that they encourage voters to think strategically. In plurality voting some voters must decide whether to vote for their true favorite, or for a slightly lower choice who they believe is more likely to win. In approval voting, voters must guess whether their goals are more likely to be achieved by approving just a few candidates, or approving more. These are burdens on the voter, and can distort results in favor of the candidates whose voters are best or luckiest at thinking strategically.

The Borda system is also susceptible to strategic voting. Suppose a voter's top two sincere choices are A and B respectively, but the voter guesses that B is the biggest threat to A's victory. The voter may then choose to put B at the very bottom, to minimize the chance B will beat A. The Borda system is also not Condorcet-consistent; I join many electoral theorists in considering Condorcet-consistency to be absolutely indispensable. The Black system (Black, 1958) picks a Condorcet winner if there is one, and otherwise picks the Borda winner. Thus it is still susceptible to the Borda strategic-voting problem.

Green-Armytage (2014) compared 8 electoral systems on 5 forms of strategic manipulability. He found that all 8 were somewhat manipulable, but Borda was worst while minimax was one of the less manipulable systems.

**4.4 Ties in Copeland**

Section 1.7 explained why I don't believe Copeland satisfies the minimum-change criterion. Another major problem with Copeland is the number of ties it produces in the event of a Condorcet paradox. To study this issue, I used random numbers as voter ratings of candidates, to generate 10,000 Condorcet paradoxes for each number of candidates from 3 to 10, with 75 voters in each case. When the number of candidates was 3 or 4, Copeland produced ties in every single trial with a Condorcet paradox. With 5 to 10 candidates, the number of such trials with Copeland ties was respectively 8656, 7565, 6650, 5966, 5490, and 5041 – in each case out of 10,000 trials exhibiting a Condorcet paradox. Thus with any number of candidates from 3 to 10, it appears that most elections exhibiting Condorcet paradoxes will have ties under the Copeland system. This rate of ties is unacceptable. Any tie-breaker added to Copeland would be used so often that it would in effect produce an entirely different system.

# 5. Simulations Comparing Electoral Systems

**5.1 Why we need computer simulations**

Section 2 described many incompatibilities between electoral system properties, all of which may seem desirable or even essential. Scholars have debated these issues for well over 200 years, and the number of incompatibilities and paradoxes has only grown. It thus seems unlikely that a purely analytic solution to these issues will appear. This suggest that the best approach may be to identify models of voter behavior which appear to be consistent with real-world data, and see which electoral systems behave best in artificial data generated by those models. Section 1.4 has suggested a family of such models. We now consider computer simulations comparing minimax-T2 to as many as 9 other electoral systems in models of this type. I did not originally create these simulations to support minimax. I was agnostic when I ran my first simulations, but redesigned the simulations to focus on minimax when it consistently emerged as best in the earlier studies. Minimax-T2 was used because the simulations used full ranking. As seen in Section 3.2, minimax-T1 breaks no ties under full ranking, and minimax-T3 is a combination of T1 and T2.

**5.2 Overview and procedures**

Section 5 describes 11 computer simulation studies, all using full ranking. Each study uses some way of identifying one computer-generated candidate as the "best" candidate, and then compares minimax-T2 to as many as 7 other well-known electoral systems (plus the new systems SSMD and SSSMD described in Section 3.3) on the frequency with which they select that candidate. These studies use the model of voter behavior introduced in Section 1.4, though they include no categorical variables.

Except in Study 11, I also omitted any term for "excellence" or "general attractiveness," for two reasons. First, the work of Tideman and Plassmann (2012) suggests this dimension is not important in real elections. Second, my own unpublished studies show that in spatial models with an attractiveness term, the Condorcet paradox arises even less often than in models with no such term. This effect is not surprising, because in both ordinary language and these models, attractiveness is a transitive concept. The Condorcet paradox occurs only when winning is intransitive (as when A beats B and B beats C but C beats A), so the more strongly voters are influenced by transitive properties like general attractiveness, the less often the paradox will appear. But I was most interested in comparing minimax to other

Condorcet-consistent electoral systems, because I agree with most other electoral theorists that Condorcet consistency is an essential feature of electoral systems. But Condorcet-consistent systems will pick different winners only when the Condorcet paradox occurs, and I wanted to compare their accuracy when they pick different winners. Therefore, I omitted the excellence dimension to make the paradox appear more often.

Thus these studies focus on spatial variables, sometimes adding a term for pure random error to model voter carelessness and misinformation. For the spatial variables I always used bivariate normal distributions with mutually independent dimensions. Those distributions were used to draw both voters and candidates; both were always drawn anew for each trial.

In Studies 1-10, five criteria were used to label one candidate as the "best" candidate. The five criteria were as follows.

*"Error" studies.* The 3 studies of this type rely on a fact mentioned in Section 1.5: if random errors are added to voter ratings of candidates in a spatial model, a set of ratings which had not exhibited a Condorcet paradox may show one. This models the effects of voter carelessness and misinformation. Samples were drawn, and mutually independent normally distributed random errors were added to all the ratings in a sample, until a sample which had not displayed a paradox later displayed one. Then 10 electoral systems were applied to the sample data, and systems were compared on the frequency with which they picked the original Condorcet winner.

*"Sampling" studies.* The two studies of this type rely on the fact, also mentioned in Section 1.5, that when a population displays no Condorcet paradox, random samples from the population may still display one. Populations may be of finite size. In each trial a population of 200 voters was generated, and a random sample of 75 voters was drawn from it. When the computer found a population which displayed a Condorcet winner, and the sample from it displayed no such winner, 10 electoral systems were applied to the sample, and systems were compared on the frequency with which they picked the candidate who had been the Condorcet winner in the population.

*"Centrism" studies.* In the 3 studies of this type, 8 electoral systems were compared on their ability to pick the candidate closest to the center of the distribution of voters – the most "centrist" or "moderate" candidate. In these studies the Condorcet paradox is produced by the sampling error in random sampling. But here I'm classifying studies by the rule used to determine the "best" candidate, so these 3 studies fall in a separate category than the "sampling" studies.

In the 8 studies of the three types just mentioned, the Schulze (2011) and Coombs (1964) electoral systems were the only previously-published systems which ever picked the "best" candidate even 70% as often as minimax-T2 did. I felt these three types of study were perhaps the most important studies in Section 5, so I eliminated all the other systems from the remaining three studies (Studies 9-11). I also eliminated the Coombs system from these studies because that system is not Condorcet consistent. I also eliminated the SSMD and SSSMD systems from these three studies, and from the centrism studies, because they were unpublished systems I had invented myself, and they had performed substantially worse than minimax-T2 in the error and sampling studies. Thus Studies 9 and 10 compared minimax-T2 only to Schulze. Study 11 is described more fully later. I now continue my listing of study types.

*An "asymmetry" study*. In Study 9 I studied situations in which a Condorcet winner was found when voter and candidate opinion scores in a two-dimensional space were drawn from a symmetric bivariate normal distribution, but a Condorcet paradox appeared when these scores were exponentiated so the modified scores were drawn from an asymmetric log-normal distribution. Using a spatial model, on the average my computer had to generate 250 trials to find one trial meeting these conditions. When a trial met these conditions, voting patterns computed from the modified scores were fed into the minimax-T2 and Schulze systems, which were then compared on the frequency with which they picked the Condorcet winner from the original symmetric scores.

*An "opinion change" study*. In each trial of Study 10 I started with a sample of 75 voters and 10 candidates, which had a Condorcet winner. Then I randomly picked one voter who had placed the Condorcet winner first, and moved that winner to last place for that one voter. A trial was counted only if this change produced a Condorcet paradox. In those trials, a Condorcet winner could be produced by changing the voting pattern of just one voter. I tested minimax-T2 and Schulze on their ability to find this hidden winner.

Study 11 used a different measure of the "best" candidate, which is described more fully later.

People may differ on the relative importance of the five study types described above (error, sampling, centrism, asymmetry, and opinion change), but in the end that matters little, because minimax-T emerged as the best system in all these studies. Readers skeptical about spatial models should note that the sampling and opinion change studies used spatial models to generate the raw data, but spatial models played absolutely no role in selecting "true winners" or in assessing the relative performance of the electoral systems being compared. Thus in the most important senses, those studies did not rely on spatial models.

Results of Study 1 appear in this paragraph, those of Studies 2-8 in Section 5.3, those of Studies 9 and 10 in Section 5.4, and those of Study 11 in Section 5.5. Studies 1-8 are organized as shown in Table 4. Study 1 examined the effect of changing the number of voters or number of candidates in "error" studies. That seemed to have little effect; minimax-T2 won every one of its $4 \times 3 \times 9$ or 108 comparisons in Study 1 (that's 4 numbers of candidates × 3 numbers of voters × 9 competitors to minimax-T2). Therefore, Studies 2-8 all used 75 voters in each trial, and all trials used 10 candidates except those involving the Kemeny system. Kemeny's computational complexity explodes with increasing numbers of candidates, so those comparisons all used just 4 candidates. Study 1 used 25,000 trials for each comparison, while Studies 2-8 all used 100,000 trials. All this is shown in Table 4. Results of Studies 2-8 appear in Tables 5-7.

As explained above, in the "sampling" and "error" studies, only trials displaying a Condorcet paradox were tabulated. When comparing two Condorcet-consistent systems on "centrism," it makes sense to follow the same rule, since we know in advance that the two systems will pick the same winner on trials without the paradox, and we're looking for differences between systems. But when comparing minimax-T2 to systems lacking Condorcet consistency, it seems to me to make more sense to compare them on all trials, since they may pick different winners on any trial. Therefore Study 5, which examines that case, is the only study which examines all trials, not just those containing a Condorcet paradox. That fact appears in the heading of Table 4, and in the "Trial type" column of Table 6.

**Table 4. Overview of 8 simulation studies comparing 9 electoral systems to minimax-T2**
All studies except #5 used only trials with a Condorcet paradox. All studies except #1 included 75 voters and 100,000 trials.

| | System compared to Minimax-T2 | | |
|---|---|---|---|
| | **Kemeny.** 4 candidates | **Other Condorcet-consistent systems: SSSMD, SSMD, Schulze, Copeland** | **Non-Condorcet-consistent systems: Coombs, Hare, plurality, approval** |
| **Study type** | | 10 candidates, except in Study 1 | |
| **Error** | **Study 6**. | **Study 1**. 5, 10, 15, or 20 candidates; 75, 175, or 275 voters. 25,000 trials for each combination. | |
| | | **Study 2** | |
| **Sampling** | **Study 7**. | **Study 3** | |
| **Centrism** | **Study 8**. | **Study 4** | **Study 5** |

Each electoral system's action on each trial in these studies was classified as a hit, miss, or tie. If a system produced a tie among two or more candidates, I didn't record whether one of the tied candidates was the candidate who had previously been labeled "best." A tie was simply considered not a hit, because in a real election a tie is a very undesirable outcome. I'll use the word "failures" to include both misses and ties.

Approval voting was implemented by having one-fifth of the voters approve just one candidate, another fifth approve two, and other fifths approve three, four and five candidates respectively, out of the 10 candidates.

**5.3 Results and discussion for studies 2-8 (on error, sampling and centrism)**
The 10 electoral systems in studies 2-8 cannot be compared directly to each other on numbers of hits, because hits were easier for some systems than for others. It was easier for Kemeny to find the best of four candidates than for other systems to find the best of ten. And the Hare, Coombs, approval, and plurality systems are not Condorcet consistent. Therefore, for reasons explained above, most of their trials included no Condorcet paradox. It's easier to find the correct winner in those trials than in trials with the paradox. Thus the numbers of hits for those four systems cannot be compared to hits for Schulze, SSSMD, SSMD, Kemeny, and Copeland, whose trials all included the paradox. Therefore, Table 5 expresses each system's number of hits as a percentage of the number of hits achieved by minimax-T2 on the same trials. Table 5 shows these percentages for each of the three study types in Table 4. The final column of Table 5 shows the mean of those percentages for each system. Systems are ordered in Tables 5-7 by these means. Except for Schulze versus Coombs, the numbers in the first three columns of Table 5 always fall in the same order as in the final column. More detail appears in Table 6, and still more in Table 7.

**Table 5. Number of hits each of 10 electoral systems got in 3 study types, expressed as percentages of the number of hits minimax-T2 got in the same 100,000 trials studied for that particular system**

| System | Error | Sampling | Centrism | Mean |
|---|---|---|---|---|
| **Minimax-T2** | 100 | 100 | 100 | 100 |
| **Schulze** | 89.5 | 92.5 | 92.1 | 91.4 |
| **Coombs** | 89.6 | 81.8 | 94.6 | 88.7 |
| **SSSMD** | 75.8 | 66.9 | | 71.4 |
| **SSMD** | 75.6 | 66.8 | | 71.2 |
| **Hare** | 68.9 | 49.2 | 55.6 | 57.9 |
| **Kemeny** | 68.8 | 47.4 | 45.7 | 54.0 |
| **Plurality** | 47.8 | 46.3 | 34.0 | 42.7 |
| **Approval** | 29.1 | 21.1 | 11.7 | 20.6 |
| **Copeland** | 16.4 | 3.8 | 3.9 | 8.0 |

It's striking how widely the systems varied from each other on the values in Table 5, and how far even the highest-scoring other systems fell below minimax-T2. Schulze and Coombs were the best of the other systems by these measures. But both these methods fail the minimum-change test of Section 1.2, declaring candidate D the loser when minimum-change says D is the clear winner. Schulze is also so complex computationally that it is effectively non-transparent to the average voter. The Coombs system is not Condorcet-consistent, and fails the voter-privacy test of Section 4.1. Working down Table 5, SSSMD and SSMD have no advantages over minimax-T2 which would lead us to give them a second look. Hare was in the middle of the 9 non-minimax systems. Kemeny has been a well-respected system, but did poorly here. The plurality and approval systems scored poorly. I presume this is because, as mentioned in Section 4.2, they use only a small fraction of the information about voter preferences that appears on a preferential ballot. As mentioned at the end of Section 5.2, for approval voting I used an arbitrary scheme to decide how many candidates were approved by each voter. Approval voting might have scored better in Table 5 if some alternative scheme had been used; that wasn't studied. Copeland was worst of all, clearly because it generated so many ties, as shown in Tables 6 and 7.

Ignoring the "Mean" column and the "Minimax-T2" row of Table 5, the table has 25 numerical entries. Each line of Table 6 corresponds to one of these entries. That is, each line compares minimax-T2 to one other system on the measure specified. The first two and last two columns of Table 6 need no explanation beyond their column headings. In the "Trial type" column, "CP" means the only trials tabulated were those with a Condorcet paradox, while "All" means all trials were tabulated. The next paragraph explains and discusses the remaining two columns.

Any two electoral systems will often pick the same winner. Therefore, when comparing the systems, we should focus primarily on trials in which the two systems picked different winners, and one was a hit and the other was a failure (a miss or a tie). I'll call these HF trials, for "hit-failure." The column in Table 6 headed "# of HF trials" shows the number of such trials for each line, out of a possible 100,000. The column headed "Minimax-T2 % wins" shows, for each line, the percentage of the HF trials in which minimax-T2 was the system with the hit. This column displays the overwhelming dominance of minimax-T2. A 75 in this column means that in the HF trials of that row, minimax-T2 beat its opponent three times as often as it lost – an enormous dominance. But of the 25 entries in that column, only 4 fall below that level. Those 4 are all for the Hare and Coombs systems in comparisons in which all trials displayed the Condorcet paradox, as indicated by the CP entries in column 3. But those two methods are not Condorcet-consistent, and therefore may do poorly in trials with a Condorcet winner. When all trials are used (see the "Trial type" column), the defeat rates for Hare and Coombs both rise above 75, with Hare far above.

**Table 6. Results of 7 simulation studies comparing minimax-T2 to other electoral systems on their ability to pick the "best" candidate.** See text for explanation.

| Opposing system | Study # | Trial type | Minimax-T2 % wins | # of HF trials | Minimax-T2 % ties | Opponent % ties |
|---|---|---|---|---|---|---|
| Using ratings containing random error, find the Condorcet winner observed when ratings were error-free | | | | | | |
| Schulze | 2 | CP | 76 | 6833 | 0 | 10 |
| Coombs | 2 | CP | 56 | 31942 | 0 | 0 |
| SSSMD | 2 | CP | 99 | 8397 | 0 | 22 |
| SSMD | 2 | CP | 96 | 9063 | 0 | 23 |
| Hare | 2 | CP | 66 | 33230 | 0 | 0 |
| Kemeny | 6 | CP | 78 | 16626 | 2 | 20 |
| Plurality | 2 | CP | 76 | 34613 | 0 | 19 |
| Approval | 2 | CP | 82 | 34708 | 0 | 0 |
| Copeland | 2 | CP | 95 | 31739 | 0 | 83 |
| Using sample data with no Condorcet winner, find the Condorcet winner in the population from which the sample was drawn | | | | | | |
| Schulze | 3 | CP | 82 | 5721 | 0 | 8 |
| Coombs | 3 | CP | 61 | 39090 | 0 | 0 |
| SSSMD | 3 | CP | 100 | 15882 | 0 | 34 |
| SSMD | 3 | CP | 99 | 16100 | 0 | 34 |
| Hare | 3 | CP | 75 | 49208 | 0 | 0 |
| Kemeny | 7 | CP | 97 | 36989 | 1 | 18 |
| Plurality | 3 | CP | 76 | 49868 | 0 | 4 |
| Approval | 3 | CP | 89 | 48435 | 0 | 0 |
| Copeland | 3 | CP | 99 | 46740 | 0 | 96 |
| Find the most centrist candidate | | | | | | |
| Schulze | 4 | CP | 82 | 6529 | 0 | 8 |
| Coombs | 5 | All | 76 | 8905 | 0 | 0 |
| Hare | 5 | All | 94 | 43273 | 0 | 0 |
| Kemeny | 8 | CP | 98 | 41917 | 1 | 37 |
| Plurality | 5 | All | 96 | 60830 | 0 | 0 |
| Approval | 5 | All | 98 | 78475 | 0 | 0 |
| Copeland | 4 | CP | 99 | 51930 | 0 | 96 |

Since no electoral system beats minimax-T2 in any of the three sections of Table 6, it's interesting to identify each system's worst performance versus minimax-T2. For 7 of the 9 electoral systems (all except Schulze and Coombs), the highest number in the "Minimax-T2 % wins" column was 94 or above. Those are percentages of the trials on which they lost to minimax-T2.

Table 7 gives more details for those interested. The table contains 25 3 × 3 sub-tables, each corresponding to a line in Table 6 and to a cell in Table 5. In each sub-table, the three rows pertain to hits, misses, and ties respectively in minimax-T2, and the three columns pertain to hits, misses, and ties respectively in the electoral system to which minimax-T2 is being compared. E. g., when minimax-T2 was compared to Kemeny on sampling, you can read in the Kemeny-sampling sub-table that there were 18,690 trials where minimax-T2 hit and Kemeny missed, 1022 where Kemeny hit and minimax-T2 missed, 17,252 where minimax-T2 hit and Kemeny tied, and 25 where Kemeny hit and minimax-T2 tied. The 9 numbers in each sub-table sum to 100,000, the number of trials in each of these studies. All entries in the last four columns of Table 6 were calculated from these values. Each 3 × 3 sub-table was collapsed into a 2 × 2 table by merging its last two rows, and its last two columns, and the values in the "Minimax-T2 % wins" and "# of HF trials" columns of Table 6 were computed from those 4 values.

**Table 7. Raw counts of hits, misses, and ties in 25 comparisons of minimax-T2 to other systems.**
See text for explanation. For reasons explained at the beginning of Section 5.3, these raw counts often cannot be compared directly to each other.

| System | Error | | | Sampling | | | Centrism | | |
|---|---|---|---|---|---|---|---|---|---|
| **Schulze** | 29068 | 1756 | 3457 | 43247 | 1085 | 3579 | 47928 | 1221 | 4148 |
| | 1620 | 57647 | 6452 | 1057 | 46332 | 4699 | 1160 | 41498 | 4044 |
| | 0 | 0 | 0 | 0 | 0 | 1 | 0 | 0 | 1 |
| **Coombs** | 16524 | 17757 | 0 | 24013 | 23898 | 0 | 78898 | 6745 | 0 |
| | 14185 | 51534 | 0 | 15191 | 36897 | 0 | 2160 | 12197 | 0 |
| | 0 | 0 | 0 | 1 | 0 | 0 | 0 | 0 | 0 |
| **SSSMD** | 25937 | 27 | 8317 | 32041 | 12 | 15858 | | | |
| | 53 | 51714 | 13952 | 12 | 33694 | 18427 | | | |
| | 0 | 0 | 0 | 0 | 0 | 1 | | | |
| **SSMD** | 25568 | 226 | 8487 | 31913 | 90 | 15908 | | | |
| | 350 | 50705 | 14664 | 102 | 33414 | 18572 | | | |
| | 0 | 0 | 0 | 0 | 0 | 1 | | | |
| **Hare** | 12331 | 21950 | 0 | 11137 | 36774 | 0 | 45003 | 40640 | 0 |
| | 11280 | 54439 | 0 | 12434 | 39654 | 0 | 2633 | 11724 | 0 |
| | 0 | 0 | 0 | 0 | 1 | 0 | 0 | 0 | 0 |
| **Kemeny** | 16487 | 8200 | 4700 | 30420 | 18690 | 17252 | 32861 | 21393 | 19626 |
| | 3593 | 71721 | 14798 | 1022 | 49868 | 19329 | 854 | 44892 | 16709 |
| | 133 | 993 | 606 | 25 | 351 | 303 | 44 | 337 | 272 |
| **Plurality** | 8029 | 19735 | 6517 | 10124 | 35687 | 2100 | 26996 | 58647 | 0 |
| | 8361 | 44923 | 12435 | 12081 | 37790 | 2217 | 2183 | 12174 | 0 |
| | 0 | 0 | 0 | 0 | 1 | 0 | 0 | 0 | 0 |
| **Approval** | 3429 | 30840 | 12 | 4795 | 43110 | 6 | 8586 | 77057 | 0 |
| | 6556 | 59116 | 47 | 5319 | 46763 | 6 | 1418 | 12939 | 0 |
| | 0 | 0 | 0 | 0 | 1 | 0 | 0 | 0 | 0 |
| **Copeland** | 4090 | 1361 | 28830 | 1492 | 374 | 46045 | 1727 | 405 | 51165 |
| | 1548 | 10470 | 53701 | 321 | 1793 | 49974 | 360 | 1502 | 44840 |
| | 0 | 0 | 0 | 0 | 0 | 1 | 0 | 0 | 1 |

## 5.4 Study 9, on asymmetry, and Study 10, on voter opinion change

Studies 9 and 10 compared minimax-T2 only to Schulze, because the previous studies, when considered together, had shown Schulze to be the most impressive competitor to minimax-T2. The only trials analyzed in these studies were those with a Condorcet paradox in the observed data, because the Schulze and minimax-T2 methods would always pick the same winners on other trials, and the goal was to study the differences between the two.

Study 9 tested a method's ability to handle asymmetric distributions; its procedure is described in Section 5.2. In this study, minimax-T2 had 74,469 hits on 100,000 trials while Schulze had 73,880. There were 73,646 trials on which both methods had hits. There were 823 trials where minimax-T2 hit and Schulze failed, and 234 trials where Schulze hit and minimax-T2 failed. Thus minimax-T2 beat Schulze on 78% of the 1057 trials where one method hit and the other failed.

As already mentioned, each trial of Study 10 had 75 voters and 10 candidates. If the trial had a Condorcet winner, the computer program randomly picked one voter who had placed the Condorcet winner first, and moved that winner to last place for that one voter. The trial was counted only if that move made the moved candidate no longer be a Condorcet winner, producing instead a Condorcet paradox. Each electoral system's challenge was to identify the candidate who had originally been the Condorcet winner. This task was non-trivial because with 75 voters and 10 candidates, the average candidate receives only 7.5 first-place votes, and the procedure turned one of those into a last-place vote for the hidden winner. But minimax-T2 found the hidden winner in 80,520 of the 100,000 trials, while Schulze found it in only 71,255 of those trials, thus failing 48% more often than minimax-T2. There were 69,080 trials in which both methods found the hidden winner. There were 11,440 trials where minimax-T2 succeeded and Schulze failed, and just 2175 trials where Schulze succeeded and minimax-T2 failed, so minimax-T2 beat Schulze in 84% of the 13,615 trials in which one method succeeded and the other failed.

These results at least hint that minimax-T2 surpasses all the other methods of Tables 5, 6, and 7 in the abilities tested in Studies 9 and 10, because Schulze had surpassed all other methods on an average of the three criteria discussed in these tables.

### 5.5 Study 11: adding an attractiveness term

Studies 1-10 all omitted a model term for excellence or general attractiveness. Study 11 explored whether Borda, Kemeny, plurality, approval, Hare, or Coombs might outperform minimax-T2 when this term is added to a spatial model. I studied these methods because they were all designed with an excellence dimension in mind. In this study I defined the "true winner" as the candidate with the highest mean voter rating, before those ratings were translated into within-voter ranks, and compared methods on the frequency with which they picked the "true winner." In this situation, only Borda outperformed minimax. I won't describe this study in more detail because of Borda's other disadvantages. First, Borda lacks Condorcet consistency. Second, it is one of the systems most susceptible to strategic voting. Third, when the attractiveness term was omitted, thus turning the model into a pure spatial model, Borda joined the other systems in underperforming minimax. When Tideman and Plassmann (2012) found pure spatial models to fit real data far better than other models, one of their ill-fitting models was the model on which Borda is based (p. 225). I was interested to see that an attractiveness term makes Borda look good, but I feel that fact has little practical importance.

# 6. CMO

## 6.1 Overview

CMO, short for "constrained multinomial optimization," is a new electoral system whose existence and properties help make the case for minimax and its variants, for reasons first mentioned in Section 1.2. The computational complexity of CMO is roughly proportional to the number of different voting patterns observed in the data. With full ranking the number of possible patterns is $c!$, where $c$ is the number of candidates listed on the printed ballots. With partial ranking (which allows write-ins, tied ranks, and unrated candidates) the number of possible patterns is far greater still. Thus CMO is practical only under full ranking with few listed candidates. That makes it practical in simulation studies but not in real-world elections. The simulation study on CMO in section 7 used full ranking with 4 candidates. I'll nevertheless explain CMO in a way that applies to both full and partial ranking. I do this mostly to show that certain approximations to CMO, introduced in Section 7, apply under either full or partial ranking.

In Section 1.5 we mentioned that a Condorcet paradox may appear in a random sample from a larger population even if there is no such paradox in that population. When CMO is given sample data containing a paradox, for each candidate X it searches for the population G most consistent with the observed data, subject to the constraint that X is the Condorcet winner in that population. A population is defined by the proportions of the voters in the population displaying each of the voting patterns observed in the sample, and consistency is measured by the likelihood ratio (LR) between the population and the observed data. The candidate with the highest LR is the CMO winner.

The use of LR is justified by the famous Neyman-Pearson lemma, described in Mood and Graybill (1963, pp. 292-294) and many other texts on statistical theory. This theorem states in effect that when distinguishing between two competing hypotheses, such as the hypotheses that candidates A and B are Condorcet winners in the larger population, tests using LR values provide the most accurate possible answers. Thus CMO is the optimum test.

CMO can also be described in terms of chi-square. First we find the aforementioned population G for each candidate. Then we multiply each proportion in G by the number of voters in the election, to find the number expected under hypothesis G to display each of the observed voting patterns. A chi-square value can then be computed showing how well these expected frequencies match the observed frequencies of the voting patterns. The chi-square may be either the familiar Pearson chi-square or a likelihood-ratio chi-square; the latter equals $2 \cdot \ln(1/LR)$. Thus by choosing the candidate with the highest LR we're choosing the candidate with the lowest value of LR chi-square.

## 6.2 Executing CMO - basics

To execute CMO, we start with a list of all the voting patterns observed in the election. Let $F$ (for "frequency") denote a column of numbers showing the number of voters with each of these patterns. Let $H$ denote some particular hypothesis we have about a larger population from which the observed votes are a sample. $H$ is a column of proportions, one for each voting pattern, showing the hypothesized proportion of that pattern in the population. Single entries in $F$ and $H$ are $f_i$ and $h_i$ respectively. $F$ and $H$ have the same number of entries. The entries in $F$ sum to $v$, the number of voters, and the entries in $H$ sum to 1. We'll often use the term *vector* to refer to a column of numbers like $F$ or $H$, though we will not otherwise use the concepts or notation of matrix algebra. As above, the number of candidates is $c$.

If there is a voting pattern which is not observed in the election, then any hypothesis which assigns a nonzero value of $h_i$ to that pattern will be less consistent with the data than some otherwise-similar hypothesis which omits that pattern. But we want to find for each candidate X the hypothesis most consistent with the data under the hypothesis that X is the true winner. We know in advance that this hypothesis will not give any weight to unobserved patterns, so we can confine the analysis to patterns actually observed in the election, dropping other patterns from $F$ and $H$.

The multinomial theorem (Hays, 1981) is used to compute the probability of observing any election outcome $F$ under any hypothesis $H$. For any given $F$, the $H$ maximizing that probability is the $H$ with entries proportional to $F$. We'll call that hypothesis $HP$, where the $P$ stands for "proportional." A single entry in $HP$ is $hp_i$. The likelihood ratio LR of any other $H$ is the ratio between the multinomial probability calculated from that hypothesis, and the probability from $HP$. LR can never exceed 1.

In the mathematical expressions in this paper, every multiplication is denoted by a dot like ·, and a string of two or more letters like $hp$ stands for a single value, not the product of $h$ and $p$. Some algebra associated with multinomial probabilities reveals that

$$LR = \exp\{\Sigma[f_i \cdot \ln(h_i)] - \Sigma[f_i \cdot \ln(hp_i)]\} \tag{1}$$

where $ln$ denotes natural logarithm, $exp\{Q\}$ means $e$ is raised to the power Q, and summations are over voting patterns. The next three subsections describe three ways to execute CMO. They're equivalent in the sense that they all pick the same winners.

## 6.3 Iterative CMO

Iterative computer programs proceed by successive approximation, often taking dozens or hundreds of iterations to converge to a final answer. Iterative CMO uses an iterative computer technique called constrained optimization, described in Dechter (2004), Leader (2004) and many other works. In constrained optimization, we want to find the values in some vector U which maximize or minimize some value Z, subject to one or more constraints on the values in U. For CMO, U is $H$, Z is LR, and the constraint is that some named candidate must be a Condorcet winner in the population. Thus we use constrained optimization to find the population which is maximally consistent with the electoral outcome, subject to that constraint. We do this separately for each candidate, and the CMO winner is the one whose status as a Condorcet winner is most consistent with the data as measured by LR. However, constrained optimization programs are complex, and sometimes stall out, refusing to complete the calculations even when stepwise procedures have no problems. And constrained optimization is essentially a "black box" to all but advanced mathematicians, while stepwise CMO is much easier both to execute and to understand.

## 6.4 Stepwise CMO by candidate

In this procedure, there is usually or always an upper limit, known in advance, on the number of computational steps necessary, and that limit is usually much smaller than the number of iterations in iterative CMO. Consider first the problem of finding LR for a single arbitrarily-selected candidate we'll call Target. For this description, suppose that Target lost 3 or more two-way races. Identify the opponent to whom Target lost by the largest margin, and call that candidate Opponent. Let $HT$ denote the unknown vector representing the hypothesis with maximum LR subject to the constraint that the race between Target and Opponent is a tie rather than a defeat for Target. The *T* in *HT* stands for "temporary," because *HT* may be just the first step toward finding the final *H* for candidate Target. To find *HT*, we first divide the voting patterns into three groups: those in which Target was preferred to Opponent, those in which Opponent was preferred to Target, and those expressing equal preference for Target and Opponent. Denote these groups as Win, Lose, and Tie respectively, and denote the total numbers of voters in the groups as *W*, *L*, and *T* respectively. Each entry in column *HP* is attached to one of these three groups. Then *HT* can be found by multiplying all the *HP* values in group Win by $(W+L)/(2 \cdot W)$, all proportions in group Lose by $(W+L)/(2 \cdot L)$, and leaving all proportions in group Tie unchanged. This transforms the totals for the Win and Lose groups into $(W+L)/(2 \cdot v)$, and for the Tie group into $T/v$. Thus the Win and Lose totals are equal, so the race between Target and Opponent will be tied; and the three totals sum to 1, as they must.

The claim that vector *HT* has maximum consistency with the data, subject to the constraint that the race between Target and Opponent must be a tie, is confirmed by what I call the multinomial proportionality theorem. This theorem states that when a subgroup of voting patterns is constrained to a fixed total size such as $(W+L)/(2 \cdot v)$, that subgroup's consistency with the data is maximized, within that constraint, by setting the individual pattern frequencies within the subgroup proportional to the patterns' observed frequencies. This theorem is proven in Section 8.1. The proof is just a minor modification of the proof of the well-known theorem that in the total vector *H* which has maximum consistency with *F*, the entries in *H* are proportional to those in *F*.

After finding *HT*, we can use it to compute Target's margins of victory and defeat against each of Target's opponents in that hypothesized population. Since we forced the values in *HT* to make Target's largest loss become a tie, there is a good chance that we will find that in the population specified by *HT*, Target wins or ties all races, and thus becomes the Condorcet winner. If so, we will say this hypothesis is "validated." If that happens, we rename *HT* as *H* and enter its values into (1) to find Target's LR.

To illustrate this case, consider candidate A in the example of Table 1. A loses only to C, so we examine the A-C race. A beat C in patterns 1 and 2, with 202 voters between them, and lost to C in the other four patterns, with 403 voters. Thus $W = 202$ and $L = 403$. The first 5 elements of *F* are each 101, and the last is 100. We have $HP = F/\Sigma f$. Some calculation then shows $\Sigma(f_i \cdot \ln(hp_i)) = -1084.0103$. In this example no voters gave tied ranks to A and C, so $W+L = v$. Thus the multipliers for entries in *HP* reduce to $v/(2 \cdot W)$ for group Win and to $v/(2 \cdot L)$ for group Lose. Therefore, the first two elements of *HT* are $101/(2 \cdot 202)$, the next three are each $101/(2 \cdot 403)$, and the last is $100/(2 \cdot 403)$. I don't show the details, but this vector was validated, so we rename it as *H*, and we find $\Sigma(F \cdot \ln(h)) = -1118.0427$. Thus $LR = \exp(-1084.0103 + 1118.0427) = 1.6594\text{e-}15$, which is less than 1 in 600 trillion.

If the newly-calculated *HT* was not validated, examine the already-calculated list showing all of Target's margins of victory or defeat under *HT*. Identify Target's largest loss in this list. Use the aforementioned adjustment process against this new opponent. That is, calculate new values of *W*, *L*, and *T*, and use them to find a new *HT*. Use that new *HT* to find new margins. Identify the largest of these new margins of defeat, use that race to compute still newer values of *W*, *L*, and *T*, etc.

Each time you find a new list of Target's margins of victory (positive margins) and defeat (negative margins), the value which had been most negative (i.e., the largest defeat) has now become 0. Other values which had been negative may become 0 or positive. Thus the number of negative values in the list should drop by at least one with each new *HT*. Ultimately all values in the list will become 0 or positive. That hypothesis is validated, and the LR calculated from it is also validated. Electoral theory is full of odd paradoxes, so I cannot say with absolute certainty that the number of steps in this process never exceeds the number of Target's defeats. But I did run one million trials, each starting with a Condorcet paradox, and never found that result.

We saw earlier in this subsection that each step in the stepwise process eliminates Target's largest loss while maintaining the greatest possible consistency between hypothesis and data. It thus follows that a chain of such steps would maintain the greatest possible consistency between hypothesis and data while eliminating all of Target's losses, making Target the Condorcet winner.

### 6.5 Stepwise CMO by step

This procedure is faster than either of the previous procedures. It picks the same winner, and finds the same value for that winner's LR. It takes advantage of the fact that if an *HT* vector is not validated, then the LR computed from it is still an upper bound on that candidate's true LR. For each candidate, a "step" is the calculation of a single new vector *HT* and its associated value of LR, and the attempt to validate those values. In the current procedure we first execute a single step for each candidate. Then we have found a winner even if the highest LR is the only one validated, because the calculated values of LR for all other candidates are below that highest one, and the true LR values for those other candidates are lower still. If we fail to find a winner after just one step, we execute another step for just the candidates who had un-validated LR values above the highest validated value. If this produces new validated LR values, see again whether the highest calculated LR is validated. Keep executing steps until a confirmed winner is found.

## 7. How CMO Relates to Minimax

### 7.1 Overview

Minimax has a surprisingly close relationship to CMO. This link can best be explained by introducing two other electoral systems. Single-step CMO is an approximation to CMO in which we start the "stepwise CMO by step" procedure of Section 6.5, but execute only its first step, and accept the step-1 winner as the winner, whether or not the choice was validated. A simulation in Section 7.2 found that this winner was validated on 99.994% of all trials, meaning that single-step CMO and full CMO choose the same winner on all those trials. But single-step CMO is mathematically equivalent to a vastly simpler system I've named minimax-L. And in a simulation study with 10 candidates, 75 voters, and over 5 million trials, minimax-L picked the same winner as classic minimax on every single trial in which none of the tested methods produced a tie. Thus the best estimate is that classic minimax picks the same winner as CMO on at least 99.994% of trials without ties. Section 7.2 explains why I say "at least."

### 7.2 Single-step CMO

This procedure is like stepwise CMO except that it performs no validations, and accepts the candidate with the highest single-step LR as the winner. As just mentioned, I ran a simulation to see how often a single step suffices to find a winner in stepwise CMO by step, using voter data generated by a spatial model with 4 candidates and 75 voters. It took 9,138,797 trials to generate 50,000 trials with Condorcet paradoxes. In those 50,000 trials, a single step produced a confirmed winner in all but 546 trials. Thus on average the computer had to generate 16,738 trials to find a trial in which a single step failed to produce a confirmed winner, since 9138797/546 = 16738. That works out to saying that a single step produced a confirmed winner on 99.994% of all trials. This figure is actually an understatement of the similarity between single-step CMO and full CMO, because in some fraction of the trials which go beyond step 1, the final confirmed winner will turn out to be the same candidate as the single-step winner. I did not perform a further study to see how often this happens, since in my simulation the question pertains to only 0.006% of all trials.

It might be objected that the figures in the previous paragraph are based on a study using a spatial model which may or may not represent the real world. The figure of one failure in 16,738 trials depends heavily on the fact that on average it took 183 trials to produce a single Condorcet paradox, since 9138797/50000 = 183, and failures occur only when there is a paradox. How realistic is that figure of 183? After a thorough examination of real-world elections, Gehrlein (2006, pp. 31-58) concludes that Condorcet paradoxes are indeed rare. Given all the other failures of real-world elections – candidates hide illnesses, winners break campaign promises, organizations supporting candidates hide their membership lists, etc. – the failure rate could be 100 times as high as the 0.006% figure in my simulation, and still not be a major problem.

## 7.3 Minimax-L

Minimax-L (the L stands for "likelihood") provides a much simpler way to calculate the LR values used in single-step CMO. In any form of CMO, each LR value is computed from equation 1 in Section 6.2, using two particular candidates as Target and Opponent, but with other candidates also in the race and in the calculations. As already explained, this is a multinomial LR so long as three or more candidates are in the race. But Equation 1 is a general equation, and for any pair of candidates whom we're momentarily calling Target and Opponent, it can be used to compute the *binomial* LR for the two-way race which leaves out all other candidates. To do that, we let vector $H$ be the two-cell column vector (0.5 0.5), and we let $F$ contain just the vote totals for those two candidates. We also let $HP = F/(W+L)$. Surprisingly, this binomial LR will equal the multinomial LR pertaining to the same two candidates. But by either hand or computer, computing time for the binomial LR is only a fraction of that for the multinomial LR. I'll prove the equality of these two LR values in two ways. The next two paragraphs present a conceptual proof showing that the two values "must" be equal. And Section 8.2 gives an algebraic proof that they actually are equal.

As described by Darlington (1990, p. 263), Darlington (2005), and Darlington and Hayes (2017, p.329), a composite hypothesis is a hypothesis asserting that all of several simpler hypotheses are true. For instance, if there are four candidates A, B, C, D, the hypothesis that A is a Condorcet winner in a population consists of the three hypotheses A beats B, A beats C, and A beats D in that population. Thus a hypothesis concerning a Condorcet winner is a composite hypothesis. Suppose we polled a small sample of voters a week before a four-candidate election, using preferential ballots like those to be used in the election. Suppose we found that A beat all others, and we wanted to confirm that A's victories were statistically significant. It is a well-established principle of statistics that to confirm a composite hypothesis like this, we can select the alternative simple hypothesis that is rejected most weakly. If even that hypothesis is rejected significantly, then we can conclude that the entire composite hypothesis is significantly confirmed. In the present example we would do that by choosing the opponent who had lost to A by the smallest margin, and show that A's victory was statistically significant even in that race, perhaps using a sign, binomial, or chi-square test.

Now suppose this pre-election poll showed a Condorcet paradox, and we wondered whether the paradox was significant, versus perhaps having occurred by chance. The hypothesis that the paradox is real (existing in the larger population) is also a composite hypothesis. For instance, if in some sample A beats B, B beats C, and C beats A, the hypothesis that the paradox is real implies that all those two-way differences exist in the same direction in the population. Thus we can again test the hypothesis by testing the significance of the smallest two-way difference. LR for this test can be found using the binomial version of Equation 1. Thus the binomial and multinomial LR values must be equal because they both provide likelihood-ratio tests of the same null hypothesis. Section 8.2 provides an algebraic proof of this conclusion.

Let minimax-L denote a version of minimax in which, if there is no Condorcet winner, we use the binomial version of Equation 1 to compute LR for each race. The sign test is the version of the binomial test in which the unknown proportion is hypothesized to be 0.5, so we're really talking about the sign test here. In minimax-L we find each candidate's lowest binomial LR. The candidate for whom this value is highest is the winner. We saw earlier in this subsection that the binomial LRs in this paragraph equal the multinomial LR's of CMO Step 1, under either full or partial ranking. Thus minimax-L is equivalent to CMO Step 1, but is much simpler. The difference in simplicity is noticeable even if we focus just on the calculations in CMO Step 1 which I have mentioned explicitly. But we mustn't forget the work involved in listing and managing all the observed voting patterns. With full ranking the number of possible voting patterns is $c!$, so with 10 candidates the number of possible voting patterns exceeds 3 million. The number under partial ranking is far greater still. With thousands of voters, merely counting the number of voters with each observed pattern is a major clerical task, yet that is just the first operation in single-step CMO. Thus the simplification gained by using minimax-L is great indeed.

### 7.4 Minimax-Z and minimax-Zs

In our notation, the well-known normal approximation to the sign test reduces to $z = (W\text{-}L)/\sqrt{(W+L)}$, as mentioned in Section 3.2. If we compute this $z$ for each loss, then select the largest absolute $z$ for each candidate, then select the candidate for whom this largest $z$ is smallest, we have a close approximation to minimax-L. Just how close is discussed in Section 7.5. Call this procedure minimax-Z. If we instead compute $z^2$, and use it in the same way as $z$, we will select the same winner without needing to compute any square roots, thus confining the calculations to those possible on a four-function calculator, for greater transparency. Call this procedure minimax-Zs, where the $s$ stands for "square."

### 7.5 Comparing classic minimax, minimax-P, minimax-Z, and minimax-L

I ran a simulation to see how often classic minimax, minimax-P and minimax-Z pick the same winners as minimax-L under partial ranking. Since single-step CMO is equivalent to minimax-L, and minimax-Zs is equivalent to minimax-Z, we can consider those methods to be included in this study. Minimax-P is equivalent to classic minimax under full ranking. I wanted to distinguish between those two methods, so I used data with partial ranking, unlike the simulations in Section 5. Each trial had 10 candidates and 75 computer-generated voters. Each "voter" randomly rated each candidate with an integer from 1 to 10, distributed independently and uniformly, so that each voter's ratings of any two candidates had a 10% chance of being tied. Each trial had 10·9/2 or 45 two-way races. A "participant" in any two-way race is a voter who expresses some preference between those two candidates. Across all trials, the average minimum number of participants in any of the 45 races was 61.30, and the average maximum number was 72.58. Thus the 45 races within a trial varied noticeably but not extremely in number of participants.

The only trials counted were those with a Condorcet paradox in which all four methods selected unique winners (no ties for the winning spot). On the average it took over 50,000 such trials to find even one trial in which any two of the four methods picked different winners. I set the program to run until it had found 100 trials with any disagreements at all. The program reached that point after 5,035,454 trials. Upon examining the 100 trials with some sort of disagreement, it turned out that in every case, classic minimax agreed with minimax-Z and minimax-L, while minimax-P disagreed with all three. Since minimax-L and single-step CMO had been found to agree with pure CMO on over 99.994% of all trials, classic minimax emerges as offering the best combination (among the methods tested in this study) of simplicity, transparency, and agreement with a theoretical ideal. But minimax-T is presumably better still.

## 8. Two Proofs Concerning CMO

### 8.1 Proof of the "multinomial proportionality theorem" of Section 6.4

I'll define the multinomial proportionality theorem as the theorem which says that when a vector $freqQ$ contains the frequencies in *some* of the cells in a multinomial experiment, and you want to find the vector $Q$ of probabilities for these cells which maximizes the multinomial probability of $freqQ$ subject to the constraint $\Sigma Q = h$, where $h$ is some positive constant, this is done by setting $Q = h \cdot freqQ / \Sigma freqQ$.

Consider first the standard proof that the vector $P$ which maximizes the multinomial probability of vector $freq$ is the vector in which the elements of $P$ are proportional to the elements of $freq$. We can ignore the factorials which appear in a multinomial probability, since they're constants unaffected by $P$. And maximizing the log of the multinomial probability will maximize the probability itself. Some algebra shows that this leads to maximizing $\Sigma freq \cdot \ln(P)$ subject to the constraint $\Sigma P = 1$. Using the LaGrange multiplier $\lambda$, we must find the partial derivatives of $\Sigma freq \cdot \ln(P) + \lambda(1 - \Sigma P)$ with respect to $P$ and $\lambda$. Finding them, setting them to 0, and solving for $P$, leads to the intuitively obvious conclusion that $P = freq / \Sigma freq$. Using different notation, this proof appears in many places on the internet, accessible through search terms such as "maximize multinomial probability."

Now let $Q$ denote just a portion of vector $P$ – the portion corresponding to voting patterns in which, say, Target beats Opponent. Let $freqQ$ denote the portion of $freq$ corresponding to those same cells. Let $W$ denote the number of voters who prefer Target to Opponent, $L$ the number who prefer Opponent to Target, $T$ the number who prefer the two candidates equally, and $v$ the total number of voters. We now want to maximize the multinomial probability of $freqQ$ subject to the constraint $\Sigma Q =$ $W/v$. By logic similar to that in the previous paragraph, we must now find the partial derivatives of freqQ'ln(Q) + λ($W/v$ - ΣQ) with respect to $Q$ and λ. This leads to the conclusion $Q = (W/v) \cdot$freqQ/(ΣfreqQ). Thus $Q$ is proportional to *freqQ*. This completes the proof. We perform a similar analysis for the set of voting patterns in which Opponent beats Target.

In CMO Step 1 we execute the previous paragraph once for each Target, in Step 2 we apply it twice for each Target, and in Step 3 we apply it three times for each Target, once for each Opponent of Target that we're currently considering.

## 8.2 Proof of equivalence between single-step CMO of Section 7.2 and minimax-L of Section 7.3

Simulation studies in sections 7.2 and 7.5 indicate that the single-step version of CMO agrees with the full exact version of CMO on over 99.99% of all trials. Thus the current proof shows that the much simpler minimax-L procedure has that same relation to full CMO. This proof applies to either full or partial ranking of candidates. All multiplications are denoted by dots like ·, and we'll sometimes use a two-letter code to stand for a single value. For instance, $fw_i$ is a single value, not the product of $f$ and $w_i$. Let $c$ denote the number of candidates and $v$ the number of voters. A voting pattern is a particular ranking of the candidates, such as B D A C. We presume we have a list of all the voting patterns actually observed in an election. The list need not include all possible patterns. Let $f_i$, where $f$ denotes "frequency," denote the number of voters choosing pattern $i$. Thus $\Sigma f = v$. Let $F$ denote the total list of frequencies.

Suppose we have some hypothesis about the proportion of instances in some larger population which exhibit each pattern. Let $h_i$ denote that hypothesized proportion for each pattern, and let $H$ denote the total hypothesis. We assume the hypothesis includes only patterns observed in the election, because we want to find a hypothesis maximally consistent with the observed data, and including unobserved patterns would only lower that consistency. Thus for any hypothesis we must have $\Sigma h = 1$.

Consider the multinomial probability of observed outcome $F$ if hypothesis $H$ is true. That multinomial probability contains the expression $v!/(f_1! \cdot f_2! ...)$. Let $HP$ denote the hypothesis in which the values of $h_i$ are proportional to the values of $f_i$. As is well known, that is the hypothesis maximally consistent with the observed outcome – the hypothesis which maximizes the multinomial probability of that outcome. Then the likelihood ratio LR for any hypothesis H is the multinomial probability for H, divided by the multinomial probability for HP. But H and HP both contain the same expression involving factorials given earlier in this paragraph. Thus that expression cancels out of LR. We will focus on LR, so we need not mention that expression again.

We will talk about maximizing LR, which is the same as maximizing ln(LR). We have

ln(LR) = Σ($f$·ln($h$)) - Σ($f$·ln($f/v$)).                     (2)

The first half of the right side of (2) is what's left of the logged multinomial probability for H after discarding the factorials, and the second half is the same thing for HP, so the difference between the two halves is ln(LR).

Equation 2 applies to any circumstance, and by redefining the values in the equation, it can apply to a single two-way race within an election, between two candidates we'll call Target and Opponent. For that purpose we are interested in just two types of voting patterns – those preferring

Target to Opponent, and those preferring Opponent to Target. Voters tying those two candidates are dropped out, and $f_1$ and $f_2$ are the frequencies of those two types of pattern. If $H$ specifies that these two candidates are equally popular, then $h_1$ and $h_2$ are both 0.5. That gives us the binomial LR value for this particular two-way race.

We're focusing on Target, so we let $W$ and $L$ (for win and loss) respectively denote the number of voters who put Target ahead of Opponent, and those who did the opposite. Thus (2) becomes

$$\ln(\text{binomial LR}) = W \cdot \ln(0.5) + L \cdot \ln(0.5) - W \cdot \ln(W/(W+L)) - L \cdot \ln(L/(W+L)) \qquad (3)$$

Now we remain focused on the same two candidates Target and Opponent, but we compute $\ln(\text{LR})$ for their race using CMO Step 1, which employs a multinomial approach. We now divide all the voting patterns into three groups, named Win, Lose, and Tie, according to whether Target beat, lost to, or tied Opponent in that one pattern. Let $fW_i$ be the $i$th pattern frequency in group Win, and let $fL_i$ and $fT_i$ be similar values for groups Lose and Tie. We have $\Sigma fW = W$ and $\Sigma fL = L$, where $W$ and $L$ are the same values as in (3). Similarly, define $T$ as $\Sigma fT_i$. We have $W + L + T = v$. Let $hW_i$, $hL_i$, and $hT_i$ be hypothesized proportions of all these patterns in a larger population. In CMO Step 1 we keep the hypothesized proportions of patterns in group Tie equal to their observed proportions, but we adjust the hypothesized proportions in groups Win and Lose to make $\Sigma hW = \Sigma hL$ while the sum of column $H$ remains at 1. This is accomplished by setting

$$hW_i = fW_i \cdot (W+L)/(2 \cdot W \cdot v) \quad (4) \qquad \text{and} \quad hL_i = fL_i \cdot (W+L)/(2 \cdot L \cdot v) \quad (5)$$

To see why this is so, consider (4). If there are no ties between Target and Opponent, then $W+L = v$, so (4) reduces to $hW_i = fW_i/(2 \cdot W)$. But $\Sigma fW_i = W$, so this equation implies $\Sigma hW_i = \frac{1}{2}$. That's what we want when there are no ties; we then want a hypothesis in which half of all voters favor Target over Opponent. But when there are ties, we want to adjust the $\frac{1}{2}$ figure down, multiplying it by the proportion with no ties, which is $(W+L)/v$. That's what (4) does. Similar comments apply to (5).

We can use (4), and the fact that $\Sigma fW = W$, to write

$$\Sigma(fW \cdot \ln(hW)) = \Sigma(fW \cdot \ln(fW_i \cdot (W+L)/(2 \cdot W \cdot v))) = \Sigma(fW \cdot \ln(fW_i)) + W \cdot \ln((W+L)/(2 \cdot W \cdot v)) \qquad (6)$$

Similarly we have

$$\Sigma(fL \cdot \ln(hL)) = \Sigma(fL \cdot \ln(fL_i)) + L \cdot \ln((W+L)/(2 \cdot L \cdot v)) \qquad (7)$$

In group Tie the hypothesized proportions are the observed proportions, so

$$\Sigma(fT \cdot \ln(hT)) = \Sigma(fT \cdot \ln(fT/v)) = \Sigma(fT \cdot \ln(fT)) - T \cdot \ln(v) \qquad (8)$$

The left sides of (6), (7), and (8) sum to $\Sigma(f \cdot \ln(h))$, so those equations imply

$\Sigma(f \cdot \ln(h))$

$= \Sigma(fW \cdot \ln(fW_i)) + W \cdot \ln((W+L)/(2 \cdot W \cdot v)) + \Sigma(fL \cdot \ln(fL_i)) + L \cdot \ln((W+L)/(2 \cdot L \cdot v)) + \Sigma(fT \cdot \ln(fT)) - T \cdot \ln(v)$     (9)

We can write $\Sigma(f \cdot \ln(f/v)) = \Sigma(f \cdot \ln(f)) - v \cdot \ln(v)$.     (10)

But (2) says ln(multinomial LR) equals the left side of (9) minus the left side of (10). The first term on the right side of (10) equals the sum of the first, third, and fifth terms in (9). Thus all those terms cancel out when we substitute (9) and (10) into (2). This leaves

$\ln(\text{multinomial LR}) = W \cdot \ln((W+L)/(2 \cdot W \cdot v)) + L \cdot \ln((W+L)/(2 \cdot L \cdot v)) + v \cdot \ln(v) - T \cdot \ln(v)$.     (11)

But $\ln((W+L)/(2 \cdot W \cdot v)) = \ln((W+L)/W) - \ln(2) - \ln(v)$ so

$W \cdot \ln((W+L)/(2 \cdot W \cdot v)) = W \cdot \ln((W+L)/(W)) - W \cdot \ln(2) - W \cdot \ln(v)$.     (12)

Similarly, $L \cdot \ln((W+L)/(2 \cdot L \cdot v)) = L \cdot \ln((W+L)/(L)) - L \cdot \ln(2) - L \cdot \ln(v)$.     (13)

When we substitute (12) and (13) into (11), four of the terms in the new expression group to

$v \cdot \ln(v) - T \cdot \ln(v) - W \cdot \ln(v) - L \cdot \ln(v)$. But these four terms sum to 0 because $v = T + W + L$.

Thus (11) reduces to

$\ln(\text{multinomial LR}) = W \cdot \ln((W+L)/W) + L \cdot \ln((W+L)/L) - W \cdot \ln(2) - L \cdot \ln(2)$     (14)

Each of the four terms on the right side of (14) will remain constant if we change its sign but also replace the term's logged quantity by its reciprocal, so for instance the first term becomes $-W \cdot \ln(W/(W+L))$. If we do that for all four terms, and also rearrange them, (14) becomes

$\ln(\text{multinomial LR}) = W \cdot \ln(0.5) + L \cdot \ln(0.5) - W \cdot \ln(W/(W+L)) - L \cdot \ln(L/(W+L))$     (15)

But the right sides of (15) and (3) are identical. This completes the proof.

## 9. Summary and Conclusions

The principle of minimum change says that the winner of a multi-candidate election should be the candidate who would become the Condorcet winner with the smallest number of voters changing their behavior in some way. In Section 1 it is shown that minimax is one of just three electoral systems meeting this criterion, in a set of 21 of the best-known electoral systems. Section 2 explains why this criterion should override 10 plausible-seeming electoral criteria which conflict with it; those criteria have in the past been used to dismiss minimax. Section 3 introduces minimax-T; it offers three tie-breakers which break nearly all ties and have several other advantages. Section 4 compares several electoral systems on simplicity, transparency, voter privacy, input flexibility, resistance to strategic voting, and rarity of ties. It concludes that minimax-T surpasses all other systems when all these factors are considered together – including other systems satisfying the criterion of minimum change. Section 5 describes 6 ways of identifying a "best" candidate in computer simulation studies, and shows that by all 6 of these rules, minimax-T surpasses up to 7 well-known competing systems at selecting the "best" candidate. Section 6 describes a new electoral system named CMO, which is the theoretically optimum maximum-likelihood system under reasonable assumptions. CMO is too complex computationally for real-world elections, but is practical in simple artificial-data problems. Simulation studies in Section 7 suggest that minimax-T picks the same winner as CMO in some 99.99% of all elections. All these points suggest that minimax-T is the best electoral system for most public elections.